\documentclass[aps,showpacs,preprintnumbers,amsmath, amssymb,nofootinbib]{revtex4}

\usepackage{float}
\usepackage{graphics,epsfig}
\usepackage{bm}
\usepackage{slashed}
\usepackage{graphicx}
\usepackage{amsmath}
\usepackage{amsfonts}
\usepackage{amssymb}
\usepackage{color}%
\usepackage{dcolumn}
\providecommand{\U}[1]{\protect\rule{.1in}{.1in}}

\begin{document}
\baselineskip=0.6 cm
\title{ Formation of Fermi surfaces and the appearance of  liquid phases in holographic  theories with
hyperscaling violation}
\author{Xiao-Mei Kuang}
\email{kuangxiaomei@sjtu.edu.cn} \affiliation{Department of
Physics, National Technical University of Athens, GR-15780 Athens,
Greece.}
\author{Eleftherios Papantonopoulos}
\email{lpapa@central.ntua.gr} \affiliation{Department of Physics,
National Technical University of Athens, GR-15780 Athens, Greece.}
\author{Bin Wang}
\email{wang_b@sjtu.edu.cn} \affiliation{Department of Physics and
Astronomy, Shanghai Jiao Tong University, Shanghai 200240, China.}
\author{Jian-Pin Wu}
\email{jianpinwu@gmail.com} \affiliation{Department of Physics, School of Mathematics and Physics,
Bohai University, Jinzhou, 121013, China,\\
State Key Laboratory of Theoretical Physics, Institute of Theoretical Physics, Chinese Academy of Sciences, Beijing 100190, China.}

\date{\today}
\vspace*{0.2cm}
\begin{abstract}
\baselineskip=0.6 cm
\begin{center}
{\bf Abstract}
\end{center}
We consider a holographic fermionic system in which the fermions
are interacting with a U(1) gauge field in the presence of a
dilaton field in a gravity bulk of a charged black hole with
hyperscaling violation.  Using both analytical and numerical
methods, we investigate the properties of the infrared and
ultaviolet Green's functions of the holographic fermionic system.
Studying the spectral functions of the system, we find that as the
hyperscaling violation exponent is varied, the fermionic system
possesses Fermi, non-Fermi, marginal-Fermi and log-oscillating
liquid phases. Various liquid phases of the fermionic system with
hyperscaling violation are also generated with the variation of
the fermionic mass. We also explore the properties of the flat
band and the Fermi surface of the non-relativistic fermionic fixed
point dual to the hyperscaling violation gravity.
\end{abstract}

\maketitle
\vspace*{0.2cm}
\section{Introduction}

The AdS/CFT conjecture and more generally the gauge/gravity
duality  offered a new venue of describing many physical systems
holographically. The holographic description allows the connection
of a $d$-dimensional quantum field  theory with its dual
gravitational theory that lives in $(d+1)$ dimensions
\cite{Maldacena:1997re,Gubser:2002,Witten:1998}. The dual nature
of these two theories means that a strong-coupling limit of one of
them corresponds to a weak-coupling limit of the other. This is a
powerful result since, by employing the gauge/gravity duality,
strongly coupled phenomena can be studied using dual gravitational
systems in weak coupling.

This holographic description has attracted
considerable interest for its potential
applications to study strongly coupled systems
related to condensed matter (CM) physics. One
noticeable application of the gauge/gravity
application to CM physics is the study of the
many-body system at finite charge density. The
dual description of such a system is achieved by
introducing in the gravity sector, charged
fermions probe coupled to the gauge field
 and exploring the ground state of the holographic
system \cite{0809.3402,0903.2477,0907.2694}. The   spectral
function of the holographic fermion system was analyzed to study
its Fermi surface, low energy excitations and possible types of
Fermi and non-Fermi liquids.

In this direction and in an attempt to describe
the various phases of a metallic state at low
temperatures, a dipole coupling to massless
charged fermions  was introduced
\cite{1010.3238,1012.3751}. Then by studying the
modified Dirac equation, it was found that the
boundary fermionic propagator produces a spectrum
which has vanishing spectral weight at a range of
energies around $\omega=0$ without the breaking
of any symmetry. Another interesting result was
that as the dipole coupling strength was varied,
the fluid possessed Fermi,  marginal Fermi,
non-Fermi liquid phases and an insulating phase
which shared similarities with the Mott
insulators including the dynamic formation of a
gap and spectral weight transfer. These results
had stimulated further studies
\cite{Guarrera:2011my,1103.3982,1205.6674,1210.5735,1304.7431,1108.6134}
on the behaviour of the dipole coupling. A
remarkable property was observed in
\cite{Alsup:2014uca} where it was found that
there exists a duality between zeroes and poles
in holographic systems with massless fermions and
a dipole coupling and this property was also
verified in \cite{Vanacore:2014hka}.

The AdS/CFT correspondence was initially applied
to describe holographic fermionic systems in
which the space of their gravity sector was
described by an AdS geometry. However, in many
condensed matter systems, there exist some phase
transitions governed by fixed points with
Lifshitz dynamical scaling and many
non-relativistic fixed points. Therefore, there
was a need to formulate the duality principle to
describe  quantum field theories violating
conformal invariance but keeping
scale-invariance, having as a gravity dual to a
gravity theory with a metric with
 Lifshitz scaling \cite{0808.1725,0812.0530,Pang:2009ad,1105.6335}
 \begin{eqnarray}
ds^2=-r^{2z}dt^2+\frac{dr^2}{r^2}+r^2dx_i^2~,
\end{eqnarray}
which is however  invariant under the scaling transformation
\begin{equation}
t\rightarrow\lambda^zt,\,\, x_i\rightarrow\lambda x_i,\,\,
r\rightarrow\lambda^{-1} r~. \label{scaling}
\end{equation}
This metric is characterized by a dynamical critical exponent
$z\neq1$, in which with $z=1$ we go back to the AdS metric.

This generalization of the holographic principle
from AdS spaces of the gravity sector to Lifshitz
spaces had produced interesting results. It was
showed that the  Lifshitz exponent $z$ in the
holographic fermion systems plays an important
role in the retarded Green's function
\cite{1112.5074,1201.1764}. For a specific value
of the critical exponent $z$, the Luttinger's
theorem is violated \cite{1201.3832}  and even a
dynamical gap can be generated in the presence of
a dipole coupling \cite{wu-lifshitz}.

More recently, a larger class of scaling metrics
besides the Lifshitz one was found, by including
both Abelian gauge field and a dilaton field in
the bulk theory.  These metrics with an overall
hyperscaling factor which can be considered as an
extension  of the Lifshitz metric have the form
\cite{1005.4690}
\begin{equation}\label{pureHV}
ds^2=r^{\frac{-2\theta}{d}}\left(-r^{2z}dt^2+\frac{dr^2}{r^2}+r^2dx_i^2\right)
\end{equation}
with $z$ and $\theta$ the dynamical critical exponent and the
hyperscaling violation exponent respectively \cite{Fisher}. Note
that the metric (\ref{pureHV}) transforms as $ds\rightarrow
\lambda^{\theta/d}ds$ under the transformation (\ref{scaling}).
Under this scaling the distance is not invariant with a non-zero
$\theta$ and according to the AdS/CFT correspondence this
indicates a hyperscaling violation in the dual field theory.
Thermodynamically in these theories the entropy scales as
$T^{(d-\theta)/z}$ while in theories with hyperscaling, the
entropy scales as $T^{d/z}$. This implies that the theory with hyperscaling violation brings in an
effective dimension $d_{eff}=d-\theta$. Holographic gravity theories with
hyperscaling violation were discussed in
\cite{1201.1905,1210.0540,Alishahiha:2014cwa}. It was found in
\cite{Edalati:2012tc} that theories with hyperscaling violation in
the probe approximation and without the presence of fermions
exhibit similarities with the behaviour of a Fermi liquid.

In this work we will consider a holographic fermion system dual to
a gravity bulk with hyperscaling violation and with finite charge
density. Our aim is to study in details the behaviour of infrared
(IR) and ultra violet (UV) Green's functions in a attempt to
understand the formation of Fermi surface and  the types of the
Fermi liquids present in these theories. We will also study the
possibility of generating dynamically a gap which will indicate
the presence of a Mott insulating phase in theories with
hyperscaling violation. Finally, we will explore the spectral
function of the non-relativistic fixed point by adding the
Lorentz-violating boundary condition into the bulk action.

Holographic study of the fermion system in the gravity dual with
hyperscaling violation was discussed in \cite{1303.6053}. Their
gravitational background was with neutral charge. The fermion
system in a charged background was studied in \cite{1209.3946}.
The UV Green's function in the probe fermions limit was studied
numerically and it was found that the increase
 of the Lifshitz factor $z$ and the
hyperscaling factor $\theta$ broadened and
smoothed out the sharp peak.

The work  is organized as follows. In Section
\ref{SScalingGeometry}  we review the charged black hole
background with hyperscaling factor and analyze the geometry in
the near horizon limit at zero temperature. We set up the
formalism describing the equation of motion in the fermionic
system in the bulk theory in Section \ref{secs.3}. In Section
\ref{secs.4}  we analytically investigate the low energy behaviour
and emergent quantum critical behavior of the retarded Green
function of the dual Fermi operator. In Section \ref{secs.5} we
numerically study various properties of the UV Green's function.
Then we explore the holographic non-relativistic fixed point with
Lorentz-violating boundary condition in Section \ref{secs.6}.
Finally Section \ref{secs.7} are our conclusions.

\section{The charged black holes with hyperscaling violation from Einstein-Dilaton-Maxwell theory}
\label{SScalingGeometry}

We start with the Einstein-Maxwell-Dilaton action in
(3+1)-dimensional spacetime \cite{1105.6335}
\begin{equation}\label{EMDaction}
S_g=-\frac{1}{16\pi G}\int \mathrm{d}^{4}x \sqrt{-g}\left[R
-\frac{1}{2}(\partial \phi)^2+V(\phi)-
\frac{1}{4}\left(e^{\lambda_1\phi}F^{\mu\nu}F_{\mu\nu}+e^{\lambda_2\phi}\mathcal{F}^{\mu\nu}\mathcal{F}_{\mu\nu}\right)\right]~.
\end{equation}
The action contains two $U(1)$ gauge fields
coupled to a dilaton field $\phi$. The $U(1)$
field $A$ with field strength $F_{\mu\nu}$ is
required to have  a charged black hole solution,
while the other gauge field $\mathcal{A}$ with
field strength $\mathcal{F}_{\mu\nu}$ and with
its coupling to the  dilaton field is necessary
to generate an anisotropic scaling. We can deduce
the equations of motion for all the fields from
the above action. The Einstein equation of motion
for the metric is
\begin{eqnarray}\label{EinsteinEquation}
R_{\mu\nu}-\frac{1}{2}Rg_{\mu\nu}=\frac{1}{2}\partial_{\mu}\phi\partial_{\nu}\phi-\frac{V(\phi)}{2}g_{\mu\nu}
+\frac{1}{2}\left[e^{\lambda_1\phi}(F_{\mu\rho} F_\nu^{\rho}
-\frac{g_{\mu\nu}}{4}F^{\rho\sigma}F_{\rho\sigma})
+e^{\lambda_2\phi}(\mathcal{F}_{\mu\rho}\mathcal{F}_\nu^{\rho}
-\frac{g_{\mu\nu}}{4}\mathcal{F}^{\rho\sigma}\mathcal{F}_{\rho\sigma})\right]~.
\end{eqnarray}
The equation of motion for the dilaton field is
\begin{eqnarray}\label{KG-DilatonEquation}
\nabla^2\phi=-\frac{dV(\phi)}{d\phi}
+\frac{1}{4}\left(\lambda_1e^{\lambda_1\phi}
F^{\mu\nu}F_{\mu\nu}+\lambda_2e^{\lambda_2\phi}
\mathcal{F}^{\mu\nu}\mathcal{F}_{\mu\nu}\right)~,
\end{eqnarray}
while the Maxwell equations for the gauge fields read
\begin{eqnarray}
\label{MawellEquation1}
\nabla_\mu\left(\sqrt{-g}e^{\lambda_2\phi}\mathcal{F}^{\mu\nu}\right)&=&0~,\\
\label{MawellEquation2}
\nabla_\mu\left(\sqrt{-g}e^{\lambda_1\phi}F^{\mu\nu}\right)&=&0~.
\end{eqnarray}
We will introduce a potential of the form
\begin{equation}\label{EMDpotential}
V(\phi)=V_0e^{\gamma\phi}~,
\end{equation}
which is very  helpful to generate a general Lifshitz solution
with hyperscaling violation \cite{1209.3946}. Here
$\lambda_1$,$\lambda_2$,$\gamma$ and $V_0$ are free parameters of
the theory to be determined. We consider the following ansatz for
the metric
\begin{equation}
ds_{4}^2= r^{-\theta}
\left(-r^{2z}f(r)dt^2+\frac{dr^2}{r^2f(r)}+r^2(dx^2+dy^2)\right)~.\label{BGsolution}
\end{equation}
Before we proceed with the solution we remark that the two gauge
fields appear in the action (\ref{EMDaction}) on the same footing.
To determine them with their corresponding parameters $\lambda_i$
we first decouple the gauge field $F_{\mu\nu}$ which is responsible for
the charge of the background black hole and from the field
equations (\ref{EinsteinEquation}), (\ref{KG-DilatonEquation}) and
(\ref{MawellEquation1}) we determine the gauge field
$\mathcal{F}_{\mu\nu}$, and then with the use of the Maxwell equation
(\ref{MawellEquation2}) we determine the gauge field $F_{\mu\nu}$. Then
the solutions are as follows \cite{1209.3946}
\begin{eqnarray}
&& \label{BGsolutionf}
f=1-\left(\frac{r_h}{r}\right)^{2+z-\theta}+\frac{Q^2}{r^{2(z-\theta+1)}}\left[1-\left(\frac{r_h}{r}\right)^{\theta-z}\right]~,
\\
&& \label{BGsolutionFrtmathcal}
\mathcal{F}_{rt}=\sqrt{2(z-1)(2+z-\theta)}e^{\frac{2-\theta/2}{\sqrt{2(2-\theta)(z-1-\theta/2)}}\phi_0}r^{1+z-\theta}~,
\\
&& \label{BGsolutionFrt}
F_{rt}=Q\sqrt{2(2-\theta)(z-\theta)}e^{-\sqrt{\frac{z-1+\theta/2}{2(2-\theta)}}\phi_0}r^{-(z-\theta+1)}~,
\\
&& \label{BGsolutionephi}
e^{\phi}=e^{\phi_0}r^{\sqrt{2(2-\theta)(z-1-\theta/2)}}~.
\end{eqnarray}
Here, $r_h$ is the radius of horizon  satisfying $f(r_h)=0$ and
$Q=\frac{1}{16\pi G}\int e^{\lambda_1 \phi}F_{rt}$ is the total
charge of the black hole. All the parameters in the action depend
on the  Lifshitz scaling exponent $z$ and hyperscaling violation
exponents $\theta$ and they can be written as
\begin{eqnarray}\label{parameters}
\lambda_1&=&\sqrt{\frac{2(z-1-\theta/2)}{2-\theta}}~,\nonumber\\
\lambda_2&=&-\frac{2(2-\theta/2)}{\sqrt{2(2-\theta)(z-\theta/2-1)}}~,\nonumber\\
\gamma&=&\frac{\theta}{\sqrt{2(2-\theta)(z-1-\theta/2)}}~,\nonumber\\
V_0&=&e^{\frac{-\theta\phi_0}{\sqrt{2(2-\theta)(z-1-\theta/2)}}}(z-\theta+1)(z-\theta+2)~.
\end{eqnarray}
Note that we have $z\geq 1$ and $\theta\geq 0$. Especially, the
above solution  is not valid for $\theta=2$. Also from equations
(\ref{BGsolutionFrtmathcal}) and (\ref{BGsolutionFrt}) we  obtain
\begin{eqnarray}\label{At}
&&
\mathcal{A}_t=-\slashed{\mu}r_h^{2+z-\theta}\left[1-\left(\frac{r}{r_h}\right)^{2+z-\theta}\right],
\\
&&
\label{RealAt}
A_t=\mu r_h^{\theta-z}\left[1-\left(\frac{r_h}{r}\right)^{z-\theta}\right],
\end{eqnarray}
where we have defined
\begin{eqnarray}\label{chemicalpontential}
&&
\slashed{\mu}=\frac{\sqrt{2(z-1)(2+z-\theta)}}{2+z-\theta}e^{\frac{2-\theta/2}{\sqrt{2(2-\theta)(z-1-\theta/2)}}\phi_0},
\\
&&
\label{Realchemicalpontential}
\mu=Q\sqrt{\frac{2(2-\theta)}{z-\theta}}e^{-\sqrt{\frac{z-1+\theta/2}{2(2-\theta)}}\phi_0}.
\end{eqnarray}
The Hawking temperature of the black hole is
\begin{eqnarray}\label{temperature}
T=\frac{(2+z-\theta)r_h^z}{4\pi}\left[1-\frac{(z-\theta)Q^2}{2+z-\theta}r_h^{2(\theta-z-1)}\right].
\end{eqnarray}

Before proceeding, we would like to remark on the
parameters $z$ and $\theta$. First, the
background solution given by the equations
(\ref{BGsolution})-(\ref{BGsolutionephi}) is
valid only for $z\geq 1$ and $\theta\geq 0$. The
case of $z=1$ and $\theta=0$ corresponds to AdS
geometry. Second, the condition $z-\theta\geq 0$
is required to make the chemical potential
well-defined in the dual field theory. Third, it
is easy to see that $\theta< 2$ from equation
(\ref{Realchemicalpontential}). Combining the
requirement of the null energy condition
$(-\frac{\theta}{2}+1)(-\frac{\theta}{2}+z-1)\geq0$
\cite{1209.3946}, one can have
$\theta\leq2(z-1)$. Thus, in this charged
background, the region of the parameters is
\begin{eqnarray}\label{ParameterRegion}
\left\{
\begin{array}{rl}
&0\leq \theta \leq 2(z-1) \quad {\rm for} ~~~1\leq z<2   \ ,   \\
&0\leq\theta<2 \quad {\rm for}~~~ z\geq2  \ .
\end{array}\right.
\,
\end{eqnarray}
For convenience, we make the following rescaling
\begin{eqnarray}\label{rescaling}
&& r\rightarrow r_h r~,~~~~~t\rightarrow
\frac{t}{r_h^z}~,~~~~~(x,y)\rightarrow \frac{1}{r_h}(x,y)~,
\nonumber\\
&& Q\rightarrow r_h^{(z-\theta+1)}Q~,~~~A_t\rightarrow r_h
A_t~,~~~\mathcal{A}_t\rightarrow r_h^{\theta-z-2} \mathcal{A}_t~.
\end{eqnarray}
In the following rescaling, we can set $r_h=1$. In addition, note
that $\phi_0$ is an integration constant and we will set
$\phi_0=0$ in the following. With such rescaling, the redshift
factor $f(r)$ and the gauge fields $\mathcal{A}_t$, $A_t$ can be
expressed  respectively as,
\begin{eqnarray}\label{rescalingfr}
&&
f=1-\frac{1+Q^2}{r^{z+2-\theta}}+\frac{Q^2}{r^{2(z-\theta+1)}}~,
\\
&& \label{rescalingAtmathcal}
\mathcal{A}_t=-\slashed{\mu}\left[1-r^{2+z-\theta}\right]~,
\\
&& \label{rescalingAtmathcal2}
A_t=\mu \left[1-\left(\frac{1}{r}\right)^{z-\theta}\right]~.
\end{eqnarray}
and  the dimensionless temperature has the form
\begin{eqnarray}\label{temperaturev1}
T=\frac{(2+z-\theta)}{4\pi}\left[1-\frac{(z-\theta)Q^2}{2+z-\theta}\right]~.
\end{eqnarray}
By setting
\begin{eqnarray}\label{Q-Mu}
Q=\sqrt{\frac{2+z-\theta}{z-\theta}}, ~~i.~e.,~~ \mu=\frac{\sqrt{2(2-\theta)(2+z-\theta)}}{z-\theta},
\end{eqnarray}
one can obtain the zero-temperature limit, in which the redshift factor $f(r)$ becomes
\begin{eqnarray}\label{frTzero}
f(r)|_{T=0}=1-\frac{2(z-\theta+1)}{z-\theta}\frac{1}{r^{z-\theta+2}}+\frac{z-\theta+2}{z-\theta}\frac{1}{r^{2(z-\theta+1)}}~.
\end{eqnarray}
Obviously, in the $r\rightarrow1$ limit,
\begin{eqnarray}\label{frTzerorh}
f(r)|_{T=0,r\rightarrow1}\simeq(z-\theta+1)(z-\theta+2)(r-1)^2\equiv\frac{1}{L_2^2}(r-1)^2.
\end{eqnarray}
Therefore, at the zero temperature, we obtain the near horizon
geometry $AdS_2\times \mathbb{R}^2$ with the  curvature radius
$L_2\equiv 1/\sqrt{(z-\theta+1)(z-\theta+2)}$ of $AdS_2$ to depend
explicitly on the  Lifshitz scaling exponent $z$ and hyperscaling
violation exponent $\theta$. So, near the horizon, the metric and
the gauge fields are given by
\begin{eqnarray} \label{MetricNearHorizon}
&&
ds^{2}=\frac{L_{2}^{2}}{\varsigma^{2}}(-d\tau^{2}+d\varsigma^{2})+dx^{2}+dy^2~,
\nonumber\\
&&
\mathcal{A}_{\tau}=\frac{\slashed{e}}{\varsigma}~,~~~~~A_{\tau}=\frac{e}{\varsigma}~,
\end{eqnarray}
with $\slashed{e}=\slashed{\mu}(2+z-\theta)L_2^2$ and
$e=\mu(z-\theta)L_2^2$ and we have considered the following
scaling limit
\begin{eqnarray} \label{ScalingLimitAdS2}
r-1=\epsilon \frac{L_{2}^{2}}{\varsigma}~,~~~
t=\epsilon^{-1}\tau~,
\end{eqnarray}
with $\epsilon\rightarrow 0$, $\varsigma$ and $\tau$ to be finite.

\section{The Dirac equation}
\label{secs.3}

\subsection{The Dirac equation}

To probe the geometry with hyperscaling violation, we consider the
following Dirac action  including the bulk minimal coupling
between the fermion and the gauge field
\begin{eqnarray}
\label{actionspinor} S_{D}=i\int d^4x
\sqrt{-g}\overline{\zeta}\left(\Gamma^{a}\mathcal{D}_{a} - m
\right)\zeta~,
\end{eqnarray}
where
$\mathcal{D}_{a}=\partial_{a}+\frac{1}{4}(\omega_{\mu\nu})_{a}\Gamma^{\mu\nu}-iqA_{a}$~.
From the above action, we can derive the following Dirac equation
in Fourier space
\begin{eqnarray}
\label{DiracEinFourier} (\sqrt{g^{rr}}\Gamma^{r}\partial_{r}- m)F
-i(\omega+q A_{t})\sqrt{g^{tt}}\Gamma^{t}F +i k
\sqrt{g^{xx}}\Gamma^{x}F =0~.
\end{eqnarray}
In the above equation, we have made a redefinition of $\zeta=(-g g^{rr})^{-\frac{1}{4}}\mathcal{F}$
and a Fourier transformation $\mathcal{F}=F e^{-i\omega t +ik_{i}x^{i}}$.
In addition, due to the rotational symmetry in $x-y$ plane, we have set $k_{x}=k$ and $k_{y}= 0$.
Choosing the following gamma matrices
\begin{eqnarray}
\label{GammaMatrices}
 && \Gamma^{r} = \left( \begin{array}{cc}
-\sigma^3 & 0  \\
0 & -\sigma^3
\end{array} \right), \;\;
 \Gamma^{t} = \left( \begin{array}{cc}
 i \sigma^1 & 0  \\
0 & i \sigma^1
\end{array} \right),  \;\;
\Gamma^{x} = \left( \begin{array}{cc}
-\sigma^2 & 0  \\
0 & \sigma^2
\end{array} \right),
\qquad \ldots~.
\end{eqnarray}
 the Dirac equation becomes
\begin{eqnarray} \label{DiracEF}
\left[(\partial_{r}+m\sqrt{g_{rr}}\sigma^3)
-\sqrt{\frac{g_{rr}}{g_{tt}}}(\omega+qA_{t})i\sigma^2 -(-1)^{I} k
\sqrt{\frac{g_{rr}}{g_{xx}}}\sigma^1 \right] F_{I} =0~,
\end{eqnarray}
where $I=1,2$. After splitting $F_{I}$ into $F_{I}=(\mathcal{A}_{I},\mathcal{B}_{I})^{T}$, one has
\begin{eqnarray} \label{DiracEAB1}
&& (\partial_{r}+m\sqrt{g_{rr}})\mathcal{A}_{I}
-\sqrt{\frac{g_{rr}}{g_{tt}}}(\omega+qA_{t})\mathcal{B}_{I}
-(-1)^{I}\sqrt{\frac{g_{rr}}{g_{xx}}}k \mathcal{B}_{I} =0~,
\\
&& \label{DiracEAB2} (\partial_{r}-m\sqrt{g_{rr}})\mathcal{B}_{I}
+\sqrt{\frac{g_{rr}}{g_{tt}}}(\omega+qA_{t})\mathcal{A}_{I}
-(-1)^{I}\sqrt{\frac{g_{rr}}{g_{xx}}}k \mathcal{A}_{I} =0~.
\end{eqnarray}
Defining $\xi_{I}\equiv \frac{\mathcal{A}_{I}}{\mathcal{B}_{I}}$
and $v=\sqrt{\frac{g_{rr}}{g_{tt}}}(\omega+q A_{t})$ the above
equations can be brought in the form of a flow equation of
$\xi_{I}$
\begin{eqnarray} \label{DiracEF1}
(\partial_{r}+2m\sqrt{g_{rr}}) \xi_{I} -\left[ v + (-1)^{I} k
\sqrt{\frac{g_{rr}}{g_{xx}}}  \right] - \left[ v - (-1)^{I} k
\sqrt{\frac{g_{rr}}{g_{xx}}}  \right]\xi_{I}^{2} =0~.
\end{eqnarray}
For the convenience of numerical calculation later, we can make a
transformation $r=1/u$, so that the flow equation (\ref{DiracEF1})
can be rewritten as
\begin{eqnarray} \label{DiracEF2}
\left(\sqrt{f}\partial_{u}-2m u^{\frac{\theta}{2}-1}\right) \xi_{I}
+\left[ \frac{\tilde{v}}{u} + (-1)^{I} k  \right]
+ \left[ \frac{\tilde{v}}{u} - (-1)^{I} k  \right]\xi_{I}^{2}
=0
~,
\end{eqnarray}
where we have redefined $\tilde{v}=\frac{u^z}{\sqrt{f}}(\omega+qA_t)$.

Since the IR geometry of the charged geometry with hyperscaling
violation is  $AdS_2\times \mathbb{R}^2$, we can easily derive the
boundary conditions of $\xi$ at the horizon $r=1$ for $\omega\neq
0$ \cite{0903.2477,0907.2694,1103.3982}
\begin{eqnarray} \label{BoundaryCNH}
\xi_{I}=i~.
\end{eqnarray}
For the case of $\omega=0$, we refer to
\cite{0903.2477,0907.2694}.

\subsection{Green's functions}
In the background with hyperscaling violation in the UV limit,
from equation (\ref{pureHV}) we know that $g_{rr}=r^{-\theta-2}$,
$g_{tt}=r^{2z-\theta}$ and $g_{xx}=g_{yy}=r^{2-\theta}$.
Therefore, the Dirac equation (\ref{DiracEF}) becomes
\begin{eqnarray}\label{DiracEFHV}
\left[\partial_r+mr^{-\frac{\theta}{2}-1}\sigma^3-r^{-1-z}(\omega+q\mu)i\sigma^2-(-1)^Ikr^{-2}\sigma^1\right]F_I
=0~.
\end{eqnarray}
Since we have $\theta<2$ from  equation (\ref{ParameterRegion}),
in the limit of $r\rightarrow \infty$, equation (\ref{DiracEFHV})
reduces to
\begin{eqnarray} \label{thetaB2}
\left(\partial_{r}+\frac{m}{r^{\frac{\theta}{2}+1}}\sigma^3\right)F_{I}
\approx 0~,
\end{eqnarray}
which gives the following solutions
\begin{eqnarray}
&&
\mathcal{A}_I=a_Ie^{\frac{2m}{\theta}r^{-\frac{\theta}{2}}}\simeq
a_I\left(1+\frac{2m}{\theta}r^{-\frac{\theta}{2}}+\ldots\right)~,
\
\\
&&
\mathcal{B}_I=b_Ie^{-\frac{2m}{\theta}r^{-\frac{\theta}{2}}}\simeq
b_I\left(1-\frac{2m}{\theta}r^{-\frac{\theta}{2}}+\ldots\right)~.
\end{eqnarray}
Thus, at the leading order, $F_I$ behaves as
\begin{eqnarray}\label{FIbIaI}
F_{I} \buildrel{r \to \infty}\over {\approx} b_{I}\left(
\begin{matrix} 0 \cr  1 \end{matrix}\right) +a_{I}\left(
\begin{matrix} 1 \cr  0 \end{matrix}\right)~,
\end{eqnarray}
which agrees well with the case of zero mass in
AdS or Lifshitz-AdS geometry.

In the regime of linear response, the boundary Green's functions
can be extracted by $G_{II}=\frac{a_I}{b_I}$. At the same time,
since
\begin{eqnarray}
\xi_{I}\equiv
\frac{\mathcal{A}_{I}}{\mathcal{B}_{I}}=\frac{a_I}{b_I}=G_{I}~,
\end{eqnarray}
the boundary retarded Green's functions can be expressed in term
of $\xi_I$
\begin{eqnarray} \label{GreenFBoundary}
G (\omega,k)=\left( \begin{array}{cc}
G_{1}   & 0  \\
0  & G_{2} \end{array} \right)= \lim_{r\rightarrow \infty}
\left( \begin{array}{cc}
\xi_{1}   & 0  \\
0  & \xi_{2} \end{array} \right)  \ .
\end{eqnarray}
Also, from equation (\ref{DiracEF2}), we can see that the Green
function has  the following properties
\begin{equation}\label{Gsym}
G_{1}(\omega,k;m)=G_{2}(\omega,-k;m),~~~G_{I}(\omega,k;-m)=-\frac{1}{G_{I}(\omega,-k;m)}.
\end{equation}

\section{Low energy behavior and emergent quantum critical behaviour}
\label{secs.4}

As pointed out in Section \ref{SScalingGeometry},
the extremal near horizon geometry of this
charged black hole with hyperscaling violation is
$AdS_2\times \mathbb{R}^2$. Thus, we can discuss
the retarded Green's function and some related
emergent quantum critical behavior by using the
matching method \cite{0907.2694}.

\subsection{IR Green's function}
For the near horizon geometry $AdS_2\times
\mathbb{R}^2$, in the limit of $\omega\rightarrow
0$, the Dirac equation is
\begin{eqnarray} \label{DiracEAdS2}
\varsigma \partial_{\varsigma}F_{I}-\left[m
L_{2}\sigma^{3}-(-1)^{I}kL_{2}\sigma^{1}-i\sigma^{2}qe\right]F_{I}=0~.
\end{eqnarray}
In the above equation,
we have chosen the same Gamma matrices as in equation
(\ref{GammaMatrices}) except $\Gamma^{\varsigma}=-\Gamma^{r}$ to
reflect the change between the radial coordinate $r$ and the
coordinate $\varsigma$. The above equation can be rewritten as
\begin{eqnarray}
\varsigma\partial_\varsigma F_I=UF_I~,
\end{eqnarray}
where
\begin{eqnarray}
U=
\left( \begin{array}{cc}
mL_2   & -(-1)^IkL_2-qe  \\
-(-1)^IkL_2+qe  & -mL_2 \end{array} \right)  \ .
\end{eqnarray}
Therefore,  near the $AdS_2$ boundary ($\varsigma\rightarrow 0$),
the leading behaviour  of $F_I$ is
\begin{eqnarray}
F_I=b_I^{(0)}v_{-}\varsigma^{-\nu_I(k)}+a_I^{(0)}v_{+}\varsigma^{\nu_I(k)}~,
\end{eqnarray}
where $v_{\pm}$ are real eigenvectors of $U$ and
$\pm\nu_I(k)$ are eigenvalues in the form
\begin{eqnarray}
\nu_I(k)=\sqrt{(m^2+k^2)L_2^2-q^2e^2}~.
\end{eqnarray}
Then the dimension in the IR CFT of the operator
$\mathcal{O}_{\vec{k}}$ is given by
$\delta_{k}=\frac{1}{2}+\nu_{I}(k)$, which obviously depends on
the Lifshitz dynamical critical exponent $z$ and hyperscaling
violating exponent $\theta$. We can exactly solve the Dirac
equation (\ref{DiracEAdS2}) in $AdS_2$  and obtain the retarded
Green¡¯s functions of $\mathcal{O}_{\vec{k}}$ in the dual IR CFT
as \cite{0907.2694}
\begin{eqnarray}
\mathcal{G}_I(k,\omega)=c(k)e^{-i \pi \nu_{I}(k)}\omega^{2\nu_I(k)}~
\end{eqnarray}

\subsection{The analytical expressions of the UV Green's function and the dispersion relation}
In this subsection, we will discuss the case of $\nu_I(k)$ being real.
The case of the imaginary $\nu_I(k)$ will be discussed in the next subsection.

In general, the bulk spacetime can be divided
into the inner and outer regions
\begin{eqnarray}
&& {\rm Inner}:
r-1=\omega\frac{L_2^2}{\varsigma},~~~\epsilon<\varsigma<\infty~, \
\\
&& {\rm Outer}: r-1>\omega\frac{L_2^2}{\epsilon}~,
\end{eqnarray}
where $\omega, \epsilon,
\omega\frac{L_2^2}{\epsilon}\rightarrow 0$, but
$\varsigma$ is finite. When $\varsigma\rightarrow
0$ and $\omega/\varsigma\rightarrow 0$, there is
a non-zero overlapping region between the inner
and outer regions. Thus, within this region,
$a_I$ and $b_I$ in equation (\ref{FIbIaI}) can be
expanded in terms of the power of $\omega$
\begin{eqnarray}
&& a_I=[a_I^{(0)}+\omega
a_I^{(1)}+\ldots]+[\tilde{a}_I^{(0)}+\omega
\tilde{a}_I^{(1)}+\ldots]\mathcal{G}_I(k,\omega)~, \
\\
&& b_I=[b_I^{(0)}+\omega
b_I^{(1)}+\ldots]+[\tilde{b}_I^{(0)}+\omega
\tilde{b}_I^{(1)}+\ldots]\mathcal{G}_I(k,\omega)~,
\end{eqnarray}
so that the UV Green's function can be expressed in terms of the
IR Green's function
\begin{eqnarray}
G_{I}(\omega,k) =K\frac{a_I^{(0)}+\omega
a_I^{(1)}+\mathcal{O}(\omega^2)+(\tilde{a}_I^{(0)}+\omega
\tilde{a}_I^{(1)}+\mathcal{O}(\omega^2))\mathcal{G}(k,\omega)}
{b_I^{(0)}+\omega
b_I^{(1)}+\mathcal{O}(\omega^2)+(\tilde{b}_I^{(0)}+\omega
\tilde{b}_I^{(1)}+\mathcal{O}(\omega^2))\mathcal{G}(k,\omega)}~,
\end{eqnarray}
where $K$ is a constant. Then, we have for $a_I^{(0)}\neq 0$
\begin{eqnarray}
\rm Im~G_{I}(\omega,k)\simeq K \frac{a_I^{(0)}}{b_I^{(0)}}
\left(\frac{\tilde{a}_I^{(0)}}{a_I^{(0)}}-\frac{\tilde{b}_I^{(0)}}{b_I^{(0)}}\right)c(k)\omega^{2\nu_I(k)}
\end{eqnarray}
which will bring us the spectral function near small frequency
\begin{eqnarray}\label{spectralfunction}
A(\omega,k)=\rm Im Tr [G]\propto c(k)\omega^{2\nu_I(k)}
\end{eqnarray}
with the scaling exponent $2\nu_I(k)$ dependent on the bulk exponent.
For $a_I^{(0)}= 0$, at small $\omega$ and
near the Fermi momentum $k_F$, to the leading
order, one has
\begin{eqnarray}\label{GIIvF}
G_{I}(\omega,k) \simeq
\frac{h_1}{\tilde{k}-\frac{\omega}{v_F}-h_2e^{i\gamma(k)}\omega^{2\nu_I(k)}}~,
\end{eqnarray}
where $v_F$, $h_1$ and $h_2$ depend on the UV
data and are usually determined numerically. From
the above equation, it is easy to conclude that
the dispersion relation is
\begin{eqnarray} \label{LdispersionA}
\tilde{\omega}(\tilde{k})\propto \tilde{k}^{\delta}, \quad {\rm
with} \quad \delta  = \begin{cases} \frac{1}{2 \nu_{I}(k_F)} &
\nu_{I}(k_F) < \frac{1}{2}\cr
            1 & \nu_{I}(k_F) > \frac{1}{2}
            \end{cases}~.
\end{eqnarray}
Usually, the Fermi momentum $k_F$ is determined
numerically. In the next Section, we will study
how both Lifshitz exponent $z$ and hyperscaling
exponent $\theta$ affect the Fermi surface
structure, the dispersion relation and what kind
of Fermi liquids they give.

\subsection{Log-periodicity}
Now we move on to study the case of the imaginary $\nu_I(k)$,
which usually gives a log-periodic oscillatory behaviour of the
fermionic systems. When
\begin{eqnarray}\label{kk0}
k^2<k_0^2\equiv\frac{q^2e^2}{L_2^2}-m^2~,
\end{eqnarray}
$\nu_I(k)$ is purely imaginary. Now, to the
leading order, the UV Green's function at small
$\omega$ becomes
\begin{eqnarray}\label{GIIImaginary}
G_{II}(\omega,k)\simeq\frac{a_I^{(0)}+\tilde{a}_I^{(0)}c(k)\omega^{-2i\lambda_I(k)}}{b_I^{(0)}+\tilde{b}_I^{(0)}c(k)\omega^{-2i\lambda_I(k)}}~,
\end{eqnarray}
where we have denoted $\nu_I(k)=-i\lambda_I(k)$ with
\begin{eqnarray}
\lambda_I(k)=\sqrt{q^2e^2-(m^2+k^2)L_2^2}~ .
\end{eqnarray}
It is easy to find that the Green's function
(\ref{GIIImaginary}) is log-periodic with a
period $\tau_k=\pi/\lambda_I(k)$ for the
imaginary $\nu_I(k)$. Therefore, we refer to the
region (\ref{kk0}) as the oscillatory region
\cite{0907.2694,wu-lifshitz}, which depends on
the Lifshitz exponent $z$ and hyperscaling
exponent $\theta$. Finally, we would like to
point out that for spinors, the log-periodic
oscillatory behaviour does not mean an
instability as it was discussed in
\cite{0907.2694,wu-lifshitz}.

\section{Properties of the UV Green's function}
\label{secs.5}
\begin{figure}
\center{
\includegraphics[scale=0.28]{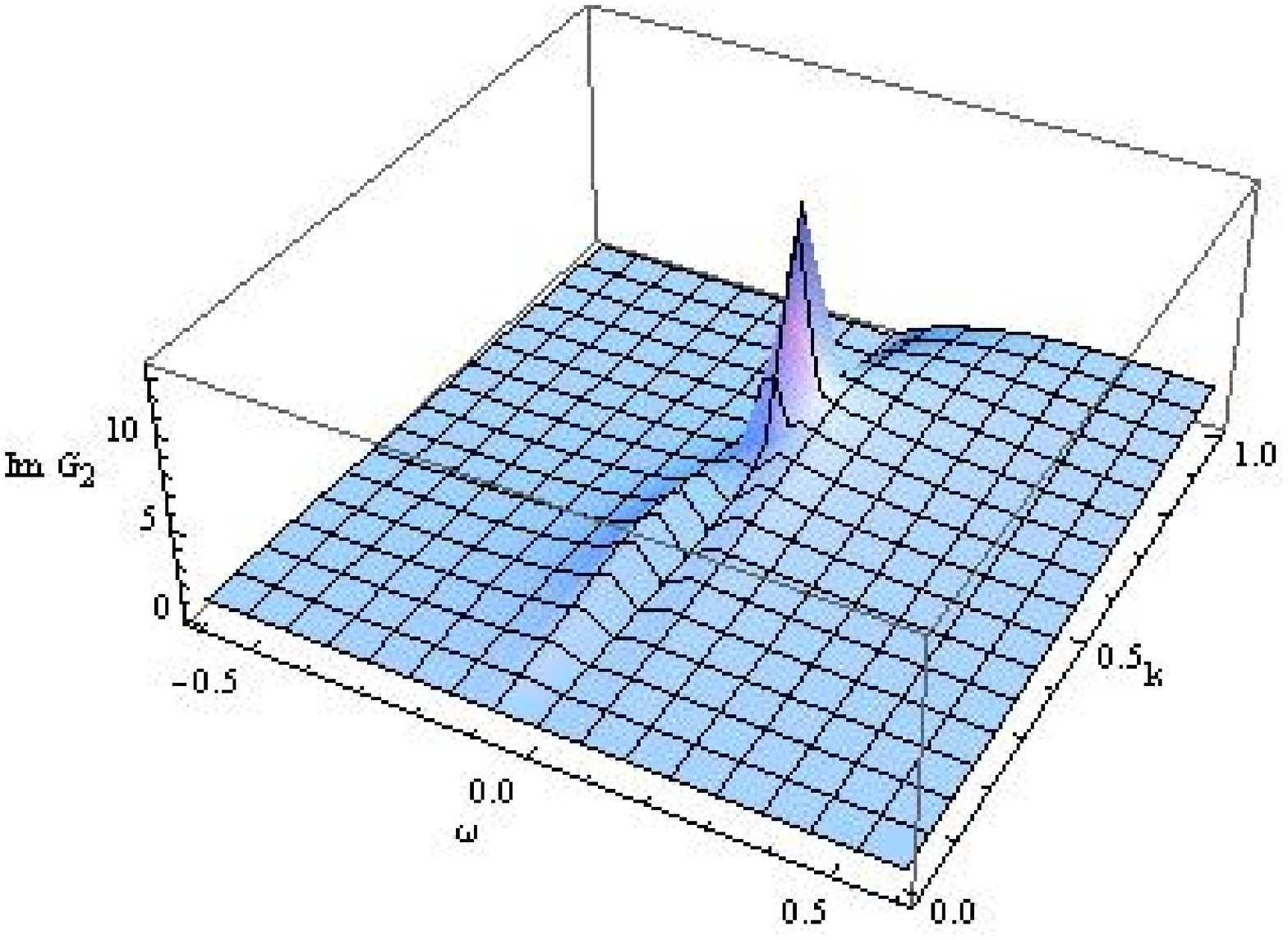}\hspace{0.5cm}
\includegraphics[scale=0.18]{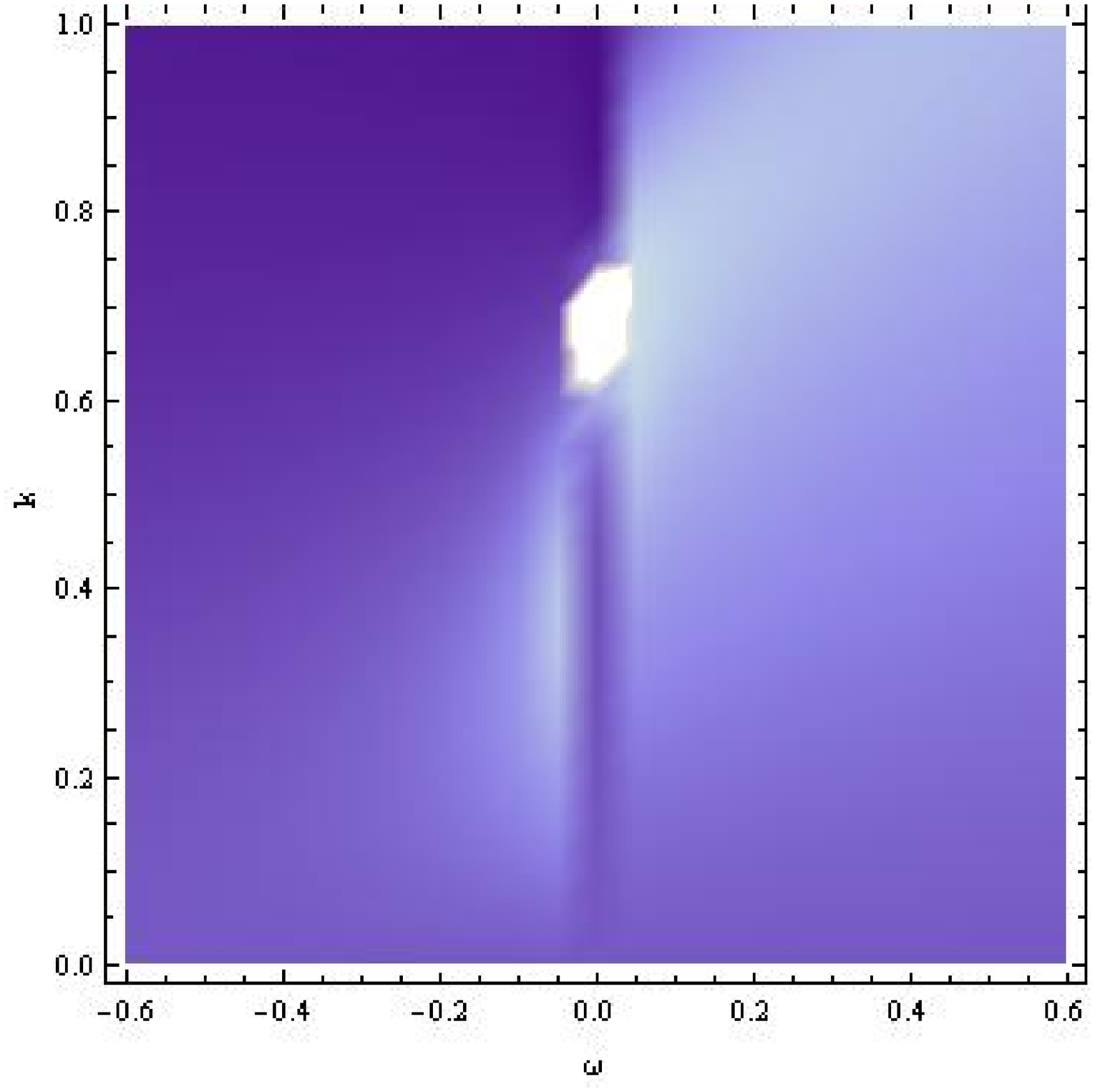}\\ \hspace{0.5cm}
\includegraphics[scale=0.28]{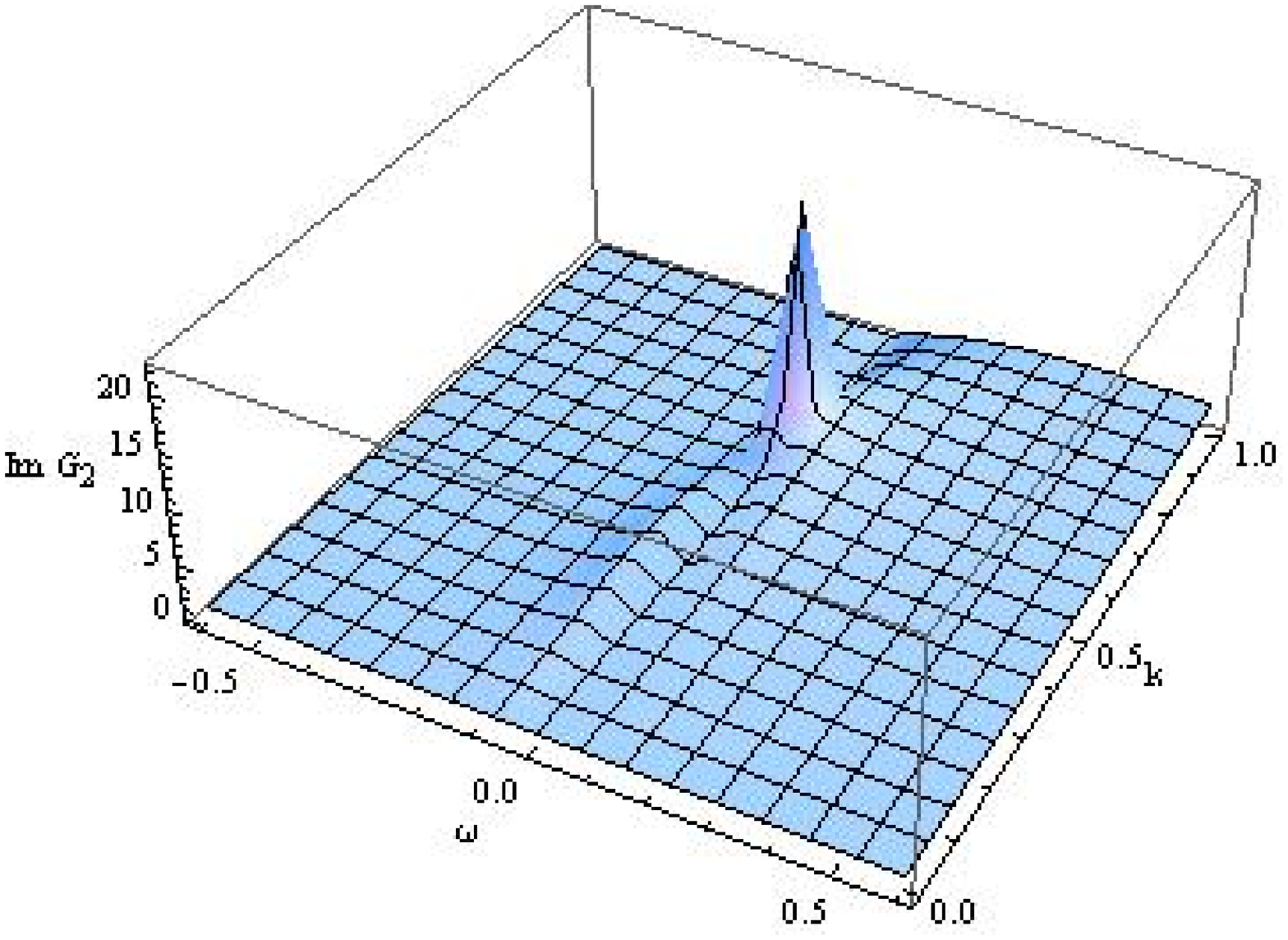}\hspace{0.5cm}
\includegraphics[scale=0.18]{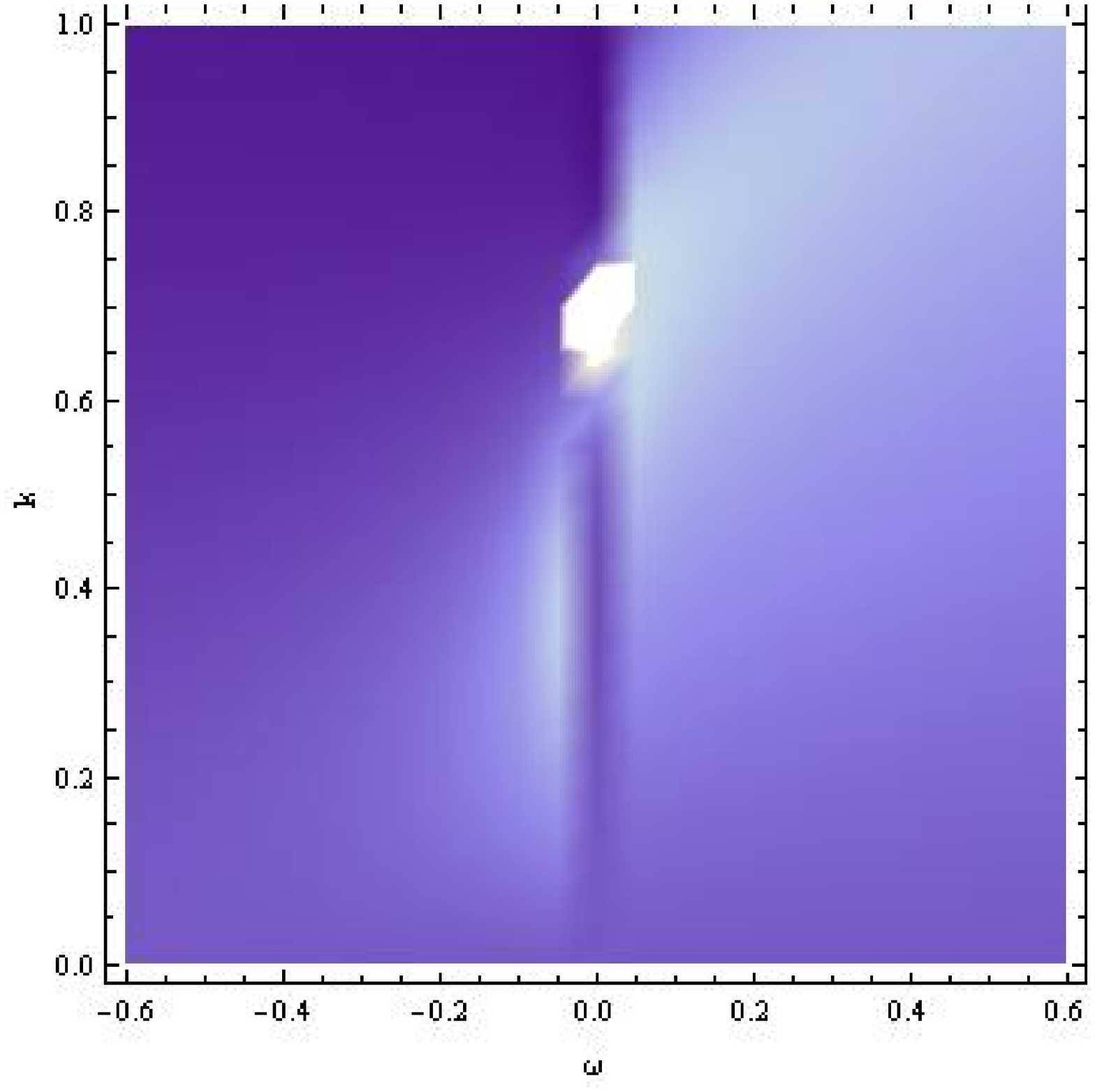}\\
\caption{\label{3dplot}The 3d and density plots
of ${\rm ImG}_{22}({\rm \omega,k})$. The plots
above are for $\theta=0$ and the plots below are
for $\theta=0.1$ ($q = 0.5$, $m=0$ and
$z=1.2$). }}
\end{figure}
\begin{figure}
\center{
\includegraphics[scale=0.6]{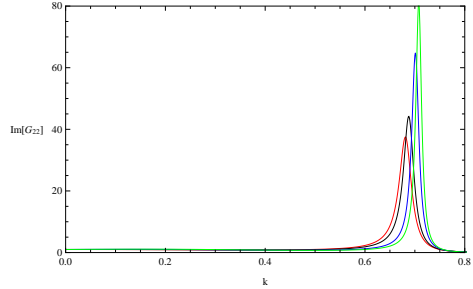}\\
\caption{\label{G22vkq05z1p02m0}The plot of ${\rm ImG}_{22}$ as a
function of $k$ for $z=1.2$ and  small $\omega$, with different
hyperscaling violating exponent $\theta$ (red for $\theta=0$,
black for $\theta=0.1$, blue for $\theta=0.3$ and green for
$\theta=0.4$). Here $q = 0.5$ and $m=0$. }}
\end{figure}

Now, we turn to  study the properties of the UV
Green's function by using numerical methods.
Previous studies on the effects of the Lifshitz
exponent $z$ to the holographic fermionic systems
showed that  the Lifshitz scaling exponent $z$
gives a clear peak in the retarded Green's
function in defining a Fermi surface and
revealing its  quasi-particle behaviour.
 Here, we will mainly focus on the effects of hyperscaling
exponent $\theta$. In Fig. \ref{3dplot}, one can notice that in
the hyperscaling violation gravity, the quasi-particle-peak
becomes wider than that in RN-AdS background as pointed out in
\cite{1209.3946}.

\subsection{Hyperscaling exponent $\theta$ dependence}
\begin{figure}
\center{
\includegraphics[scale=0.4]{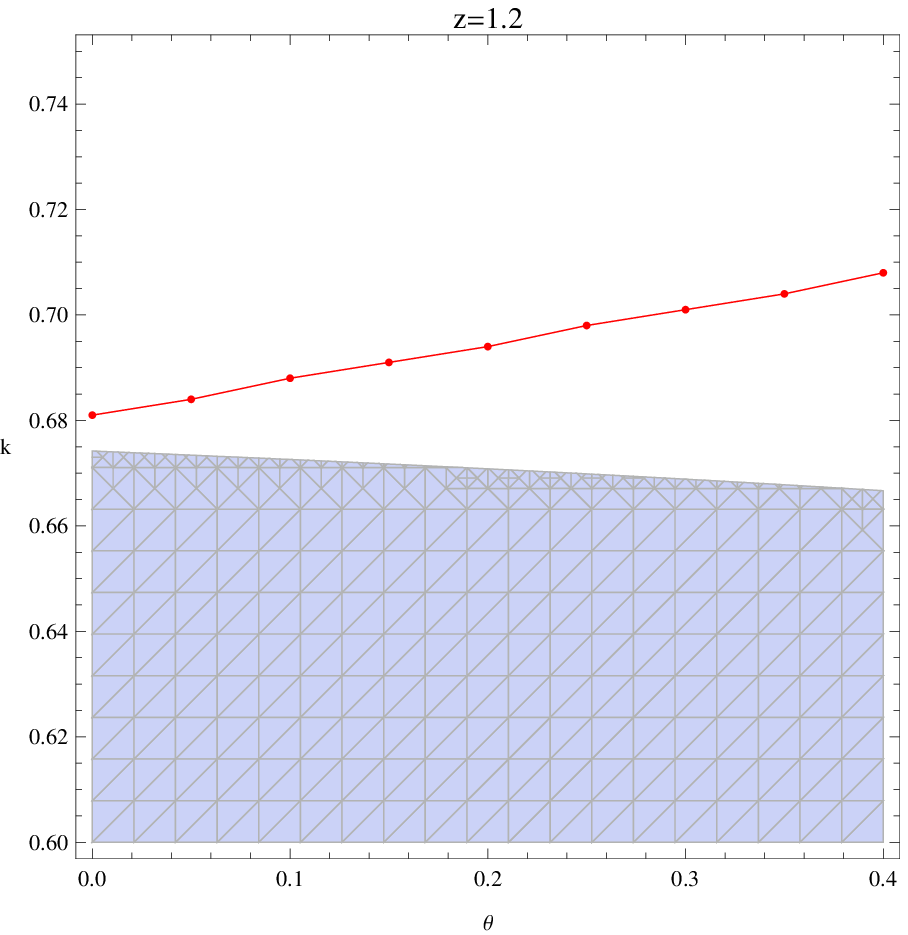}\hspace{0.5cm}
\includegraphics[scale=0.6]{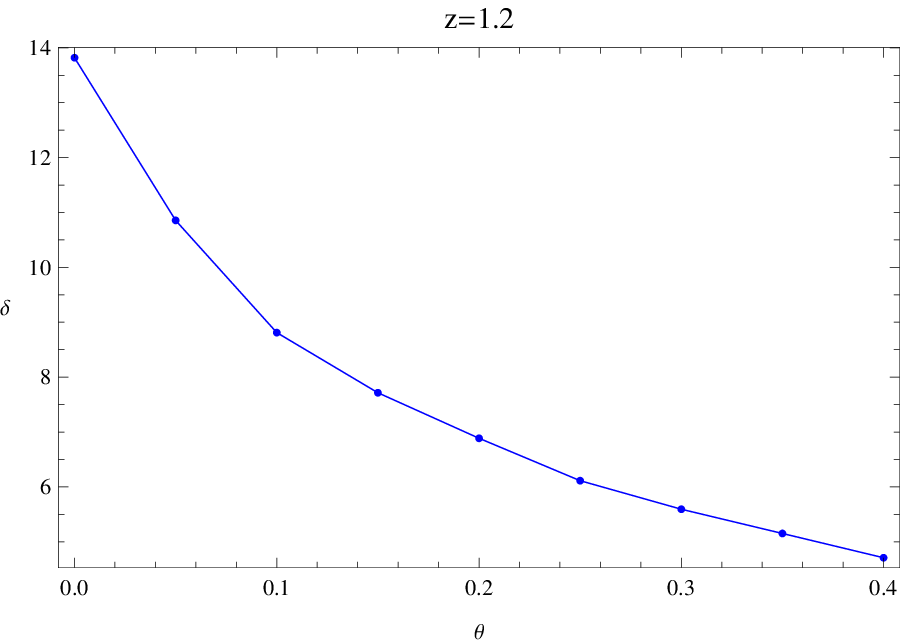}\\ \hspace{0.05cm}
\includegraphics[scale=0.4]{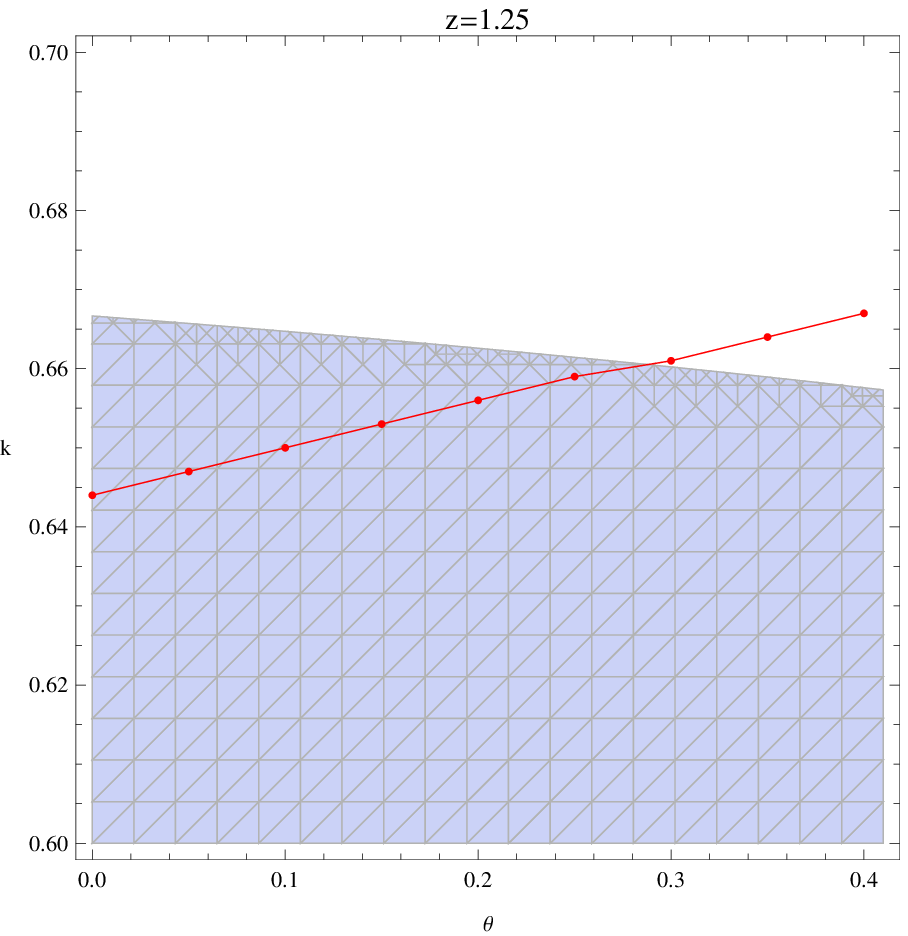}\hspace{0.5cm}
\includegraphics[scale=0.6]{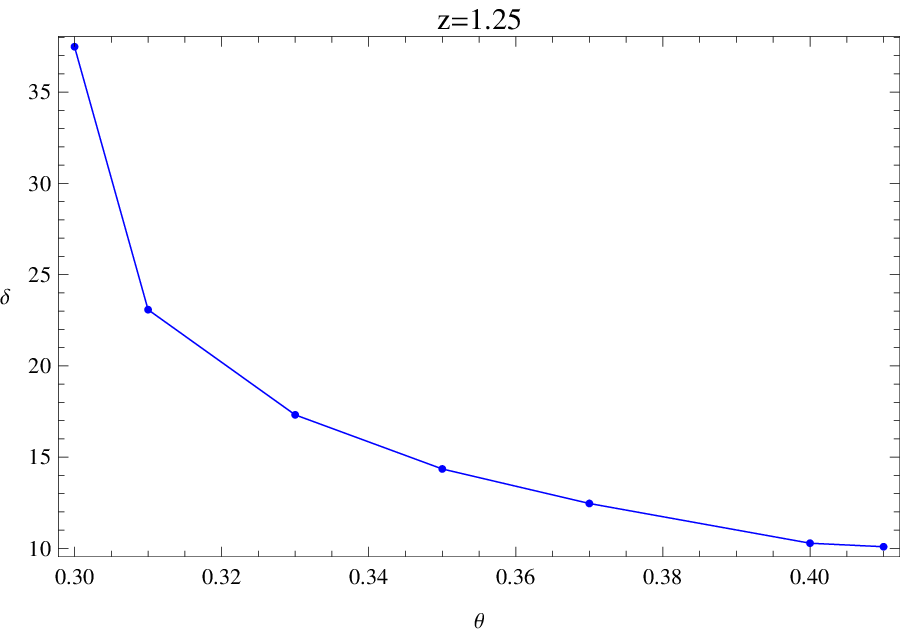}\\
\caption{\label{kFvthetam0q0p5}Left plots: The
red line is the relation between the hyperscaling
exponent $\theta$ and the location of the peak
for $\omega\rightarrow 0$ for fixed Lifshitz
exponent $z$. The blue shade in the $k-\theta$
space indicates the  unstable oscillatory region
and above it is the stable region. Right plots:
The relation between the hyperscaling exponent
$\theta$ and the scaling exponent $\delta$ of
dispersion relation for fixed Lifshtiz exponent
$z$. Here, $q=0.5$, $m=0$. In addition, $z=1.2$
in the plot above and $z=1.25$ in the plot below.
}}
\end{figure}

Now, we will study how the Fermi momentum $k_F$
and the scaling exponent $\delta$ of the
dispersion relation depend on the hyperscaling
exponent $\theta$ for fixed $q$, $z$ and zero
mass of the fermions and what types of Fermi
liquids we get depending on the value of
$\theta$.

We show the Green's function with the momentum near zero frequency
in hyperscaling violation gravity in Fig. \ref{G22vkq05z1p02m0}.
We find that  for fixed $z=1.2$, larger $\theta$ corresponds to
higher Fermi momentum, indicating that the hyperscaling exponent
$\theta$ has different effects compared to the Lifshitz scaling
exponent $z$. This can be explained by the fact that the
non-zero hyperscaling exponent introduces an effective dimension of
the theory $d_{eff}=d-\theta$. So larger $\theta$ means lower
effective dimension which calls for higher Fermi momentum. This
dependence of Fermi momentum on the dimension with minimal
coupling was first disclosed and understood in \cite{1205.6674}.

In the left plot in Fig. \ref{kFvthetam0q0p5},
the relation between the hyperscaling exponent
$\theta$ and the location of the peak for
$\omega\rightarrow 0$ is presented. In the
$k-\theta$ space the region in the blue shade is
the  log-periodic oscillatory region
\cite{0903.2477,0907.2694}. When the location of
the peak falls in the region above, it indicates
a Fermi surface. But when the location is in the
oscillatory region, the peak loses the meaning of
Fermi surfaces. From the left plot in Fig.
\ref{kFvthetam0q0p5}, we find that in the range
of allowed $\theta$, all the peaks are located in
the region above for $z=1.2$, but for $z=1.25$,
when $\theta\leq\theta_c \simeq 0.3$, the peak
begins to enter the oscillatory regime and then
it loses its meaning  as Fermi surface.
Furthermore, one can find that in the range of
allowed $\theta$, when $z\leq1.21$, all the
peaks lie outside the oscillatory region
corresponding to a Fermi liquid. But for
$z\geq1.22$, the peaks begin to enter the
oscillatory region when $\theta$ is smaller than
some critical value $\theta_c$. This indicates
the existence of a marginal Fermi liquid.

Once the Fermi momentum $k_F$ is worked out numerically, the
dispersion relation can be determined by equation
(\ref{LdispersionA}). Obviously, from the right plots in Fig.
\ref{kFvthetam0q0p5}, the exponent $\delta$ of the dispersion
relation decreases rapidly as the hyperscaling exponent $\theta$
becomes larger due to the decrease the effective dimension of the
dual theory, which is consistent with the dependence of dispersion
relation on the dimension discussed in \cite{1205.6674}. This
indicates that with the increase of $\theta$, it shows
smaller degree of deviating from the Landau Fermi liquid
phase to a non-Fermi liquid phase.

\subsection{Mass dependence}

It is well known that the types of fermion
liquids and possible transition from Fermi to
non-fermi liquids depend also on the fermion mass
in the holographic fermionic systems
\cite{0904.1993,0907.2694,0903.2596}. For
comparison, we first present the results of the
mass dependence in a fermionic system in a
Reissner-Nordstr\"om-AdS (RN-AdS) black hole
background and then we will discuss  the mass
dependence in the charged black hole with
hyperscaling violation.

\subsubsection{Mass dependence in a gravity bulk with a RN-AdS black hole}\label{MassRN}
\begin{figure}
\center{
\includegraphics[scale=0.7]{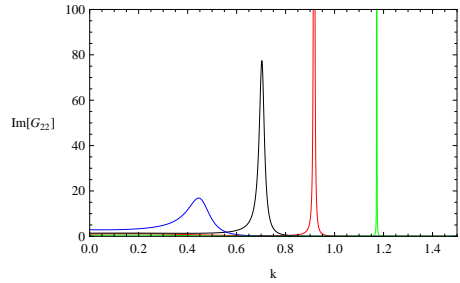}\\
\caption{\label{G22vkq05RNDm}The plot of ${\rm ImG}_{22}$ as a
function of $k$  in the RN-AdS black hole for tiny $\omega$, with
different $m$ (blue for $m=0.4$, black for $m=0.1$, red for $m=0$
and green for $m=-0.2$). Here $q = 0.5$. }}
\end{figure}
\begin{figure}
\center{
\includegraphics[scale=0.4]{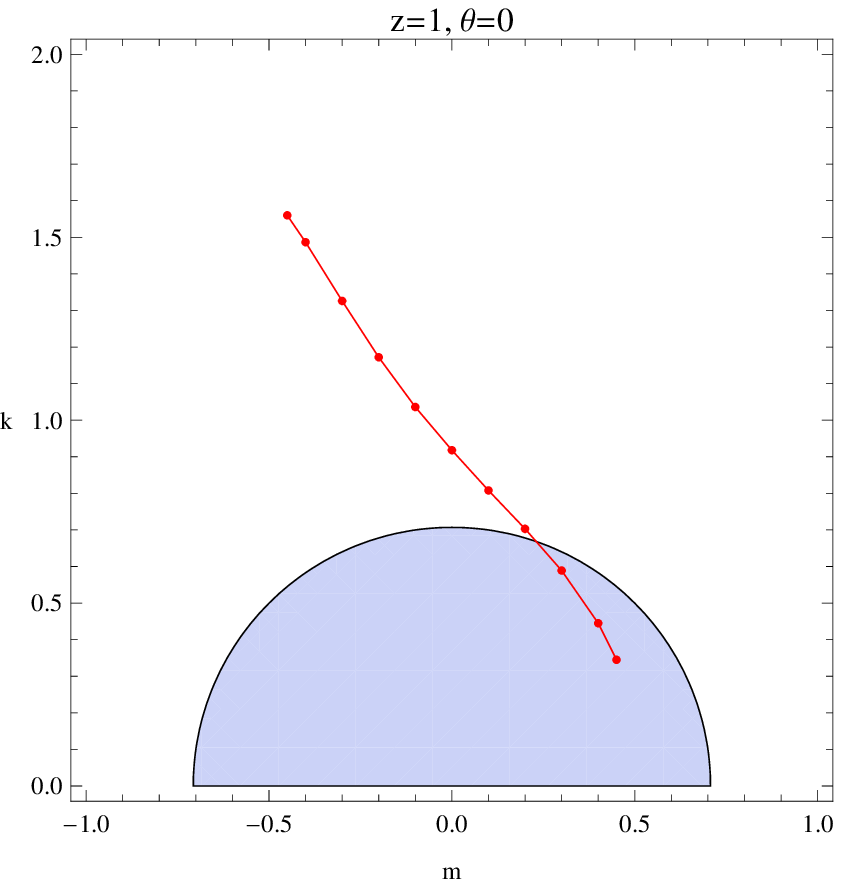}\hspace{0.5cm}
\includegraphics[scale=0.6]{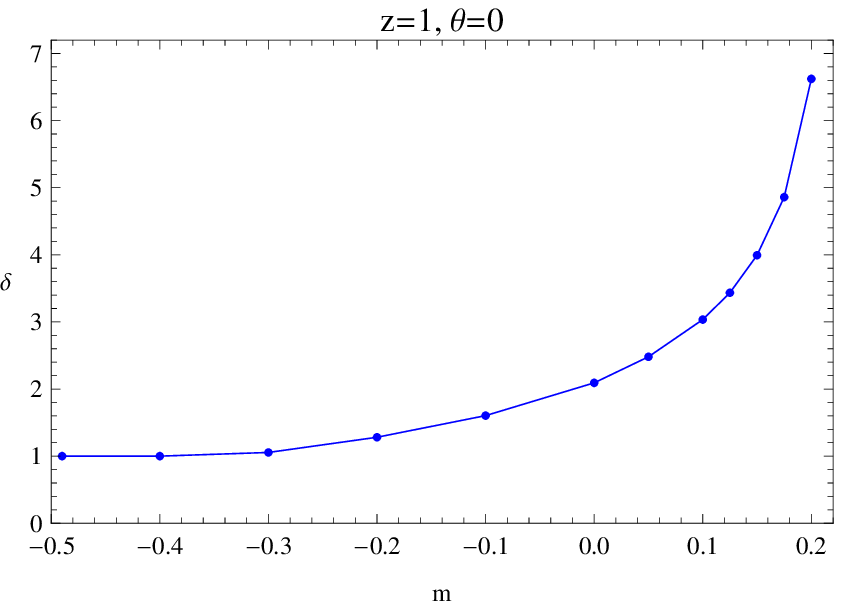}\\
\caption{\label{kFvmRN}Left plots: The relation
between the mass $m$ and the location of the peak
for $\omega\rightarrow 0$ in the RN-AdS black
hole. The shade region is the oscillatory region.
Right plots: The scaling exponent $\delta$ of
dispersion relation as a function of $m$ in the
RN-AdS black hole. Here, $q=0.5$. }}
\end{figure}
The UV Green's function in the gravity bulk with a RN-AdS black
hole is \cite{0903.2477,0907.2694}
\begin{eqnarray} \label{GreenFBoundaryRN}
G (\omega,k)= \lim_{u\rightarrow 0}u^{-2m}
\left( \begin{array}{cc}
\xi_{1}   & 0  \\
0  & \xi_{2} \end{array} \right)  \ .
\end{eqnarray}
For the purpose of numerical calculation, we make a transformation $G_{II}=u^{-2m}\xi_I$, so that one has
\begin{eqnarray} \label{DiracEF1RN}
\left(\sqrt{f}\partial_{u}+2m u^{-1}\sqrt{f}-2m u^{-1}\right)
G_{II} +\left[ \frac{\tilde{v}}{u} + (-1)^{I} k  \right]u^{-2m} +
\left[ \frac{\tilde{v}}{u} - (-1)^{I} k  \right]u^{2m}G_{II}^{2}
=0~.
\end{eqnarray}
Solving the above equation with the boundary condition (\ref{BoundaryCNH}),
one can directly read off the UV boundary Green's function $G_{II}$.

In Fig. \ref{G22vkq05RNDm}, we plot the Green's
function ${\rm ImG}_{22}$ vs. $k$ for tiny
$\omega$ with different $m$. One can
 observe that with the
decrease of $m$, the quasi-particle-peak becomes sharper and the
Fermi momentum $k_F$ is larger. Furthermore, we present the
relation between $m$ and the location of the peak for
$\omega\rightarrow 0$ in the left plot in Fig. \ref{kFvmRN}. One
finds that the Fermi momentum $k_F$ almost linearly decreases with
the increase of $m$. When $m>0.2$, the quasi-particle-peak begins
to enter into the oscillatory region and loses its meaning of
Fermi surface.

After Fermi momentum $k_F$ has been worked out
numerically,  we can use equation
(\ref{LdispersionA}) to calculate the scaling
exponent $\delta$ of the dispersion relation. The
result is presented in the right plot in Fig.
\ref{kFvmRN} where one observes that with the
decrease of $m$, the scaling exponent $\delta$
decreases. When $m\leq-0.4$, $\delta=1$, which is
a linear dispersion relation. It indicates that
there is a transition from non-Fermi liquid to
Fermi liquid as the $m$ decreases in holographic
fermionic system with a RN-AdS black hole in its
dual gravity bulk.

\subsubsection{Mass dependence in charged black hole gravity bulk with hyperscaling violation}\label{MassHV}
\begin{figure}
\center{
\includegraphics[scale=0.4]{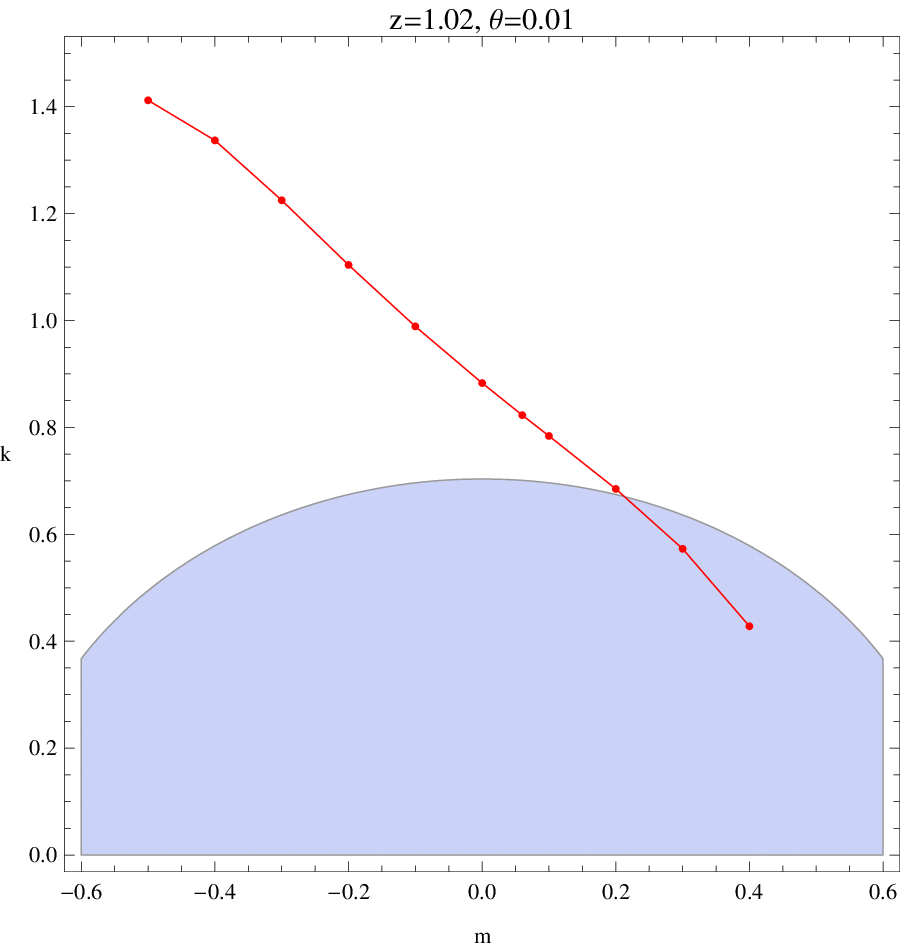}\hspace{0.5cm}
\includegraphics[scale=0.6]{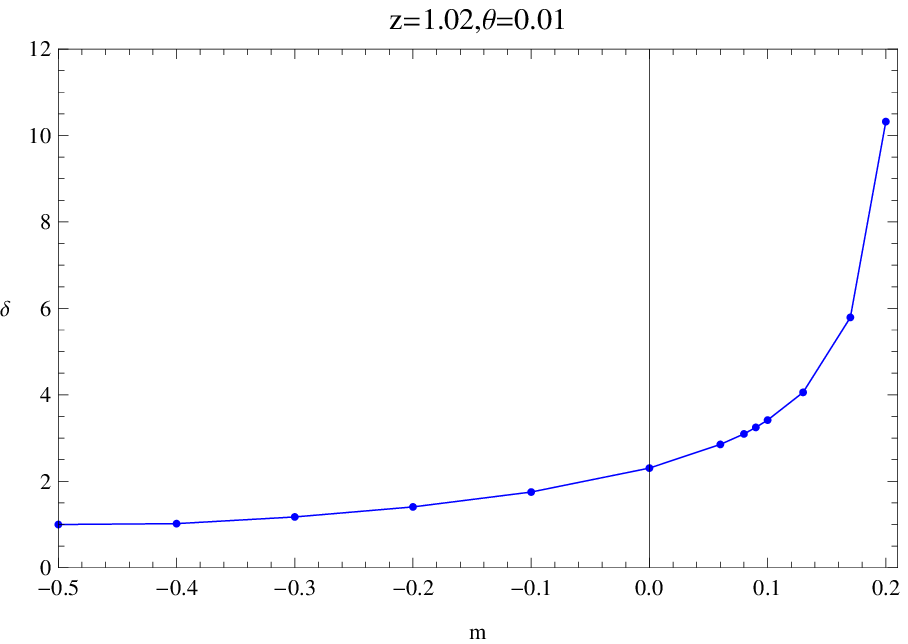}\\ \hspace{0.5cm}
\includegraphics[scale=0.4]{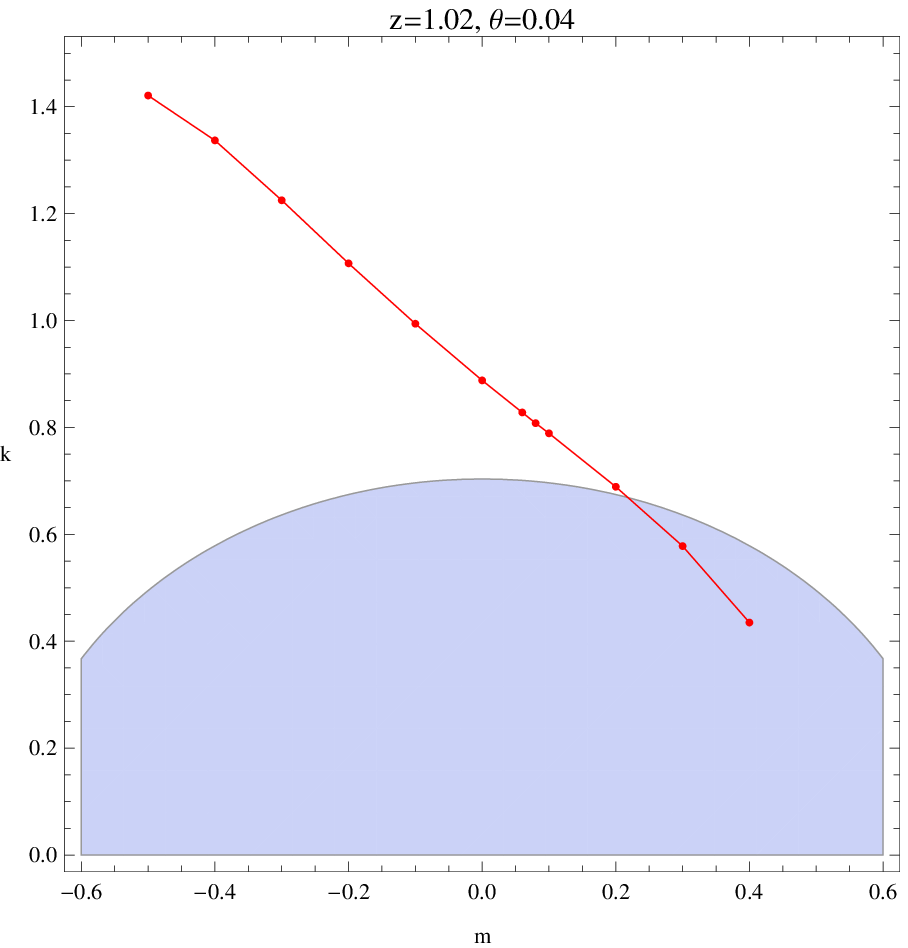}\hspace{0.5cm}
\includegraphics[scale=0.6]{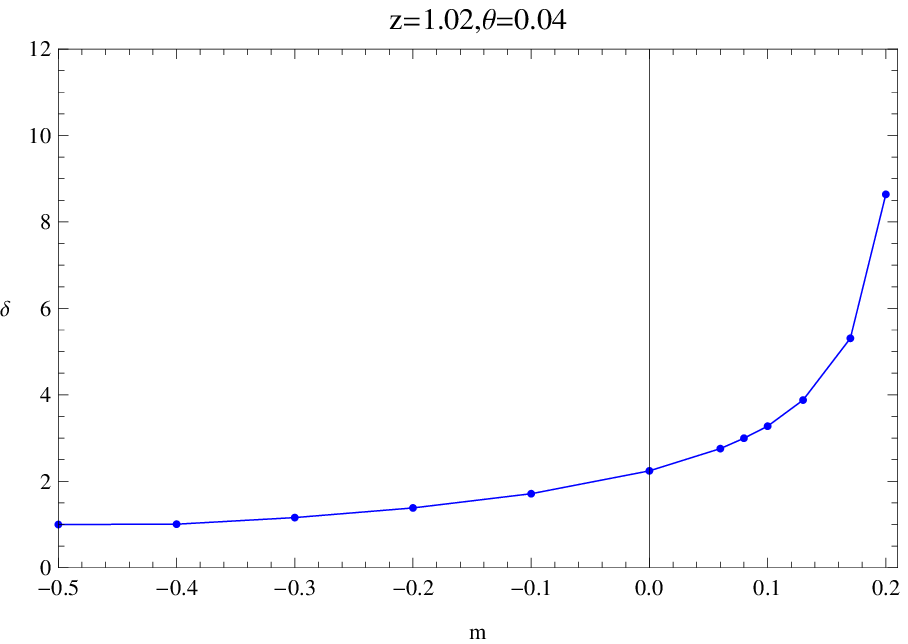}\\ \hspace{0.5cm}
\includegraphics[scale=0.4]{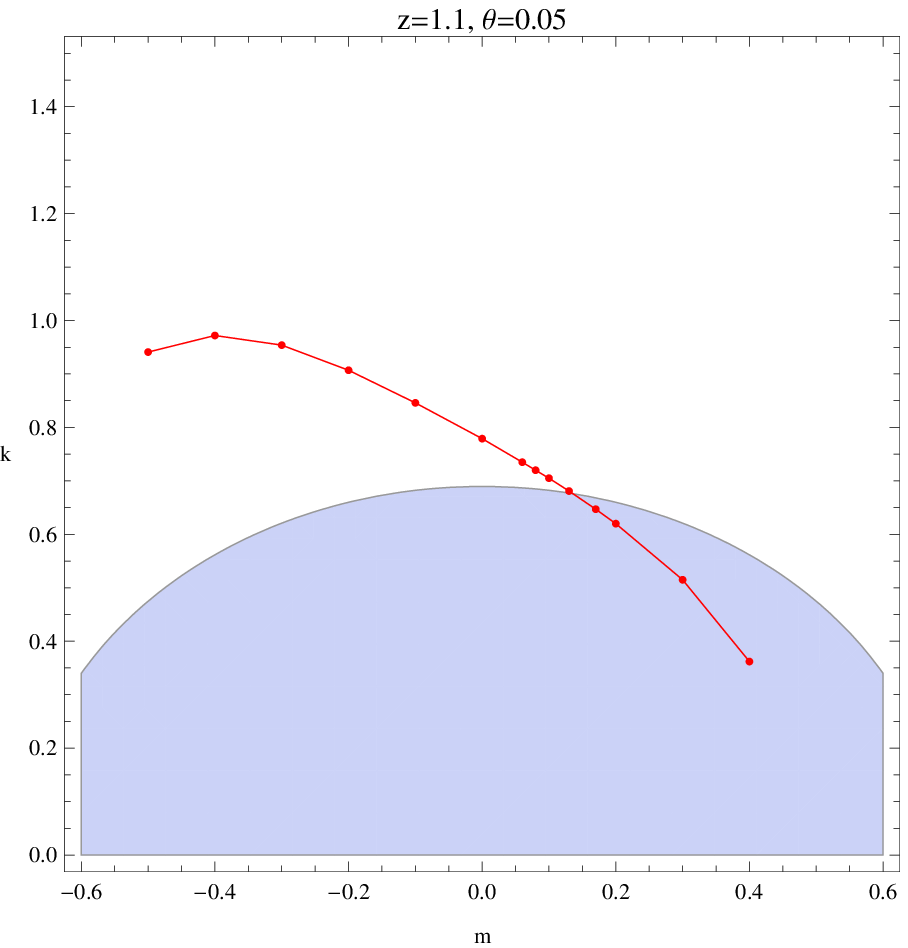}\hspace{0.5cm}
\includegraphics[scale=0.6]{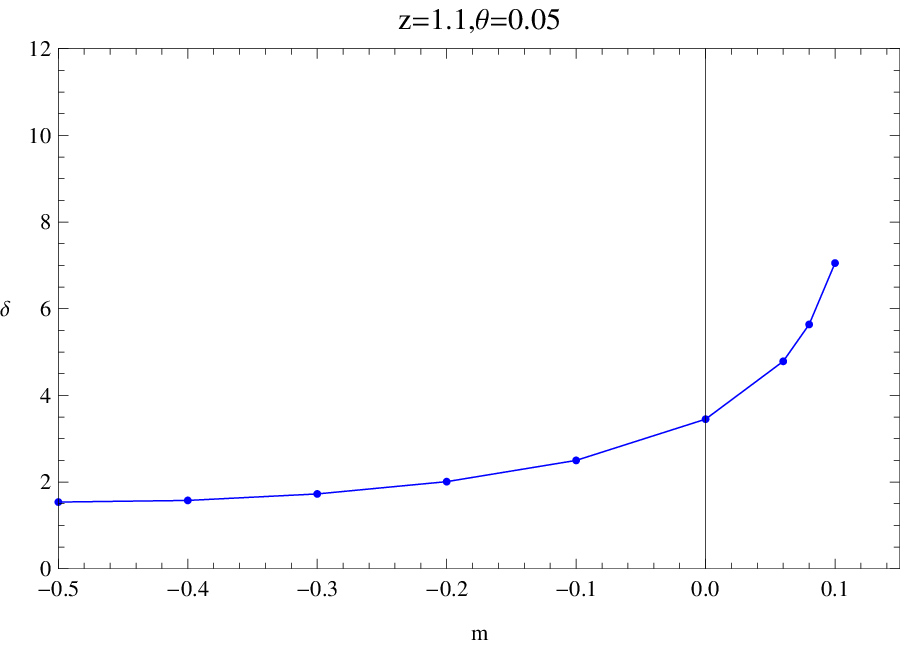}\hspace{0.5cm}\\
\caption{\label{kFvmhyperscaling}Left plots: The
relation between the mass $m$ and the location of
the peak for $\omega\rightarrow 0$ with different
$z$ and $\theta$. The shade region is the
oscillatory region. Right plots: The scaling
exponent $\delta$ of dispersion relation as a
function of $m$ with different $z$ and $\theta$.
Here, $q=0.5$. }}
\end{figure}

Now, we turn to study the mass dependence in
charged black hole gravity bulk with hyperscaling
violation. The relations between $m$ and the
location of the peak for $\omega\rightarrow 0$ as
well as the scaling exponent $\delta$ as a
function of $m$ with different $z, \theta$ are
presented in Fig. \ref{kFvmhyperscaling}. The
characteristics of the mass dependence in charged
black hole gravity bulk with hyperscaling
violation are summarized as follows
\begin{itemize}
  \item For small Lifshitz exponent $z$ and hyperscaling exponent $\theta$,
  the Fermi momentum $k_F$ almost linearly decreases with the increase
  of $m$ as that in RN-AdS black hole bulk (the above two plots in the left plots in Fig. \ref{kFvmhyperscaling}).
  In addition, with the decrease of $m$, the scaling exponent $\delta$
  decreases and there is a transition from non-Fermi liquid to Fermi liquid as the $m$ decreases
  (the above two plots in the right of Fig. \ref{kFvmhyperscaling}).
  \item From the bottom plot in the left of Fig. \ref{kFvmhyperscaling}, we can see that for large Lifshitz exponent $z$
  and hyperscaling exponent $\theta$, the Fermi momentum $k_F$ still linearly decreases with the increase of $m$ in the region of large $m$.
  However it does not persist when $m$ approaches the low bound ($m=-0.5$).
   Also, in the region of $m\in (-0.5,0.5)$, the scaling exponent $\delta$ is always larger than $1$,
  which indicates that it is a non-Fermi liquid (the bottom plot in the right of Fig. \ref{kFvmhyperscaling}).
\end{itemize}

\subsection{Failure to generate a dynamical gap in holographic
fermionic systems with hyperscaling violation}
\begin{figure}
\center{
\includegraphics[scale=0.2]{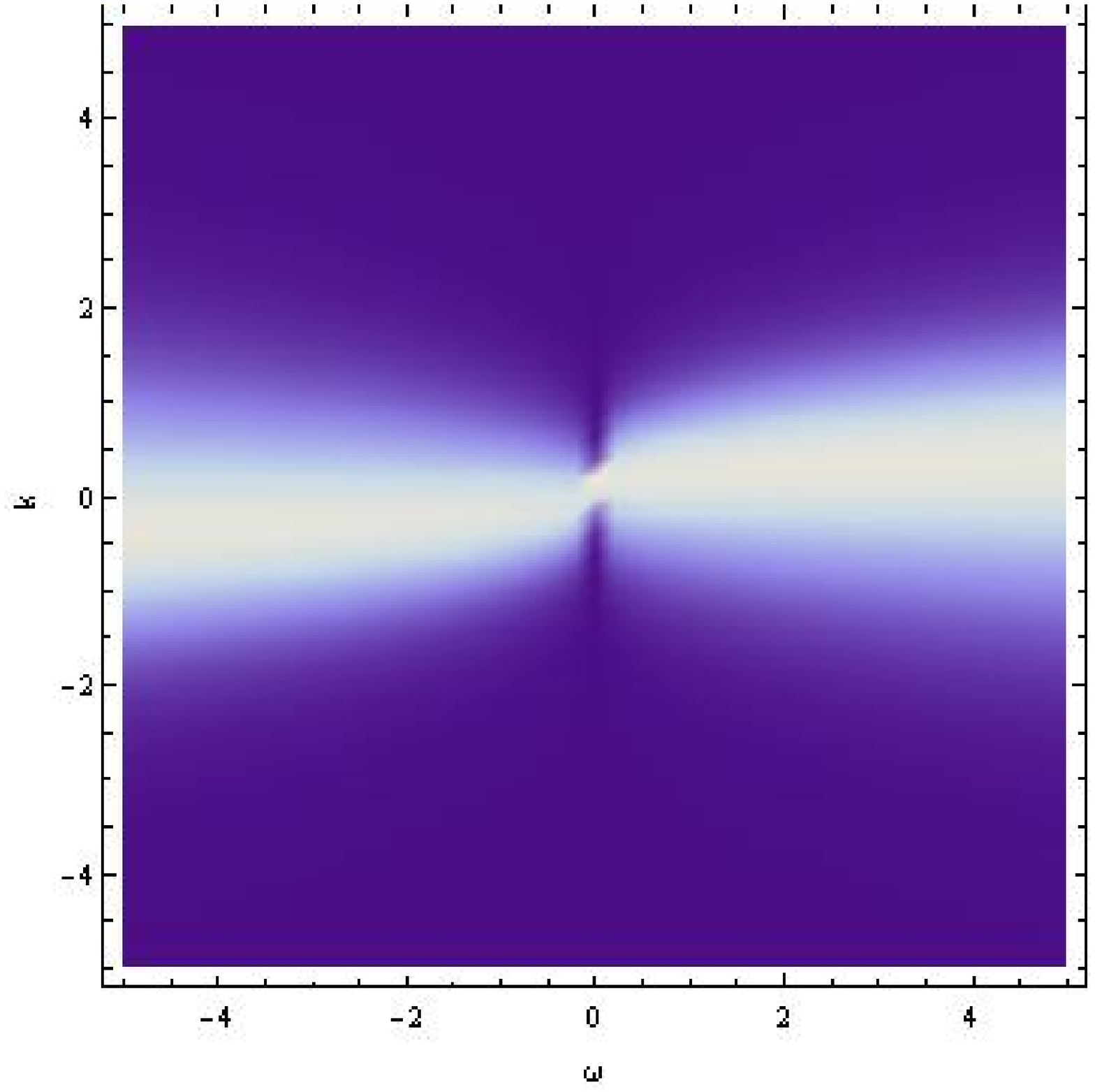}\hspace{0.5cm}
\includegraphics[scale=0.2]{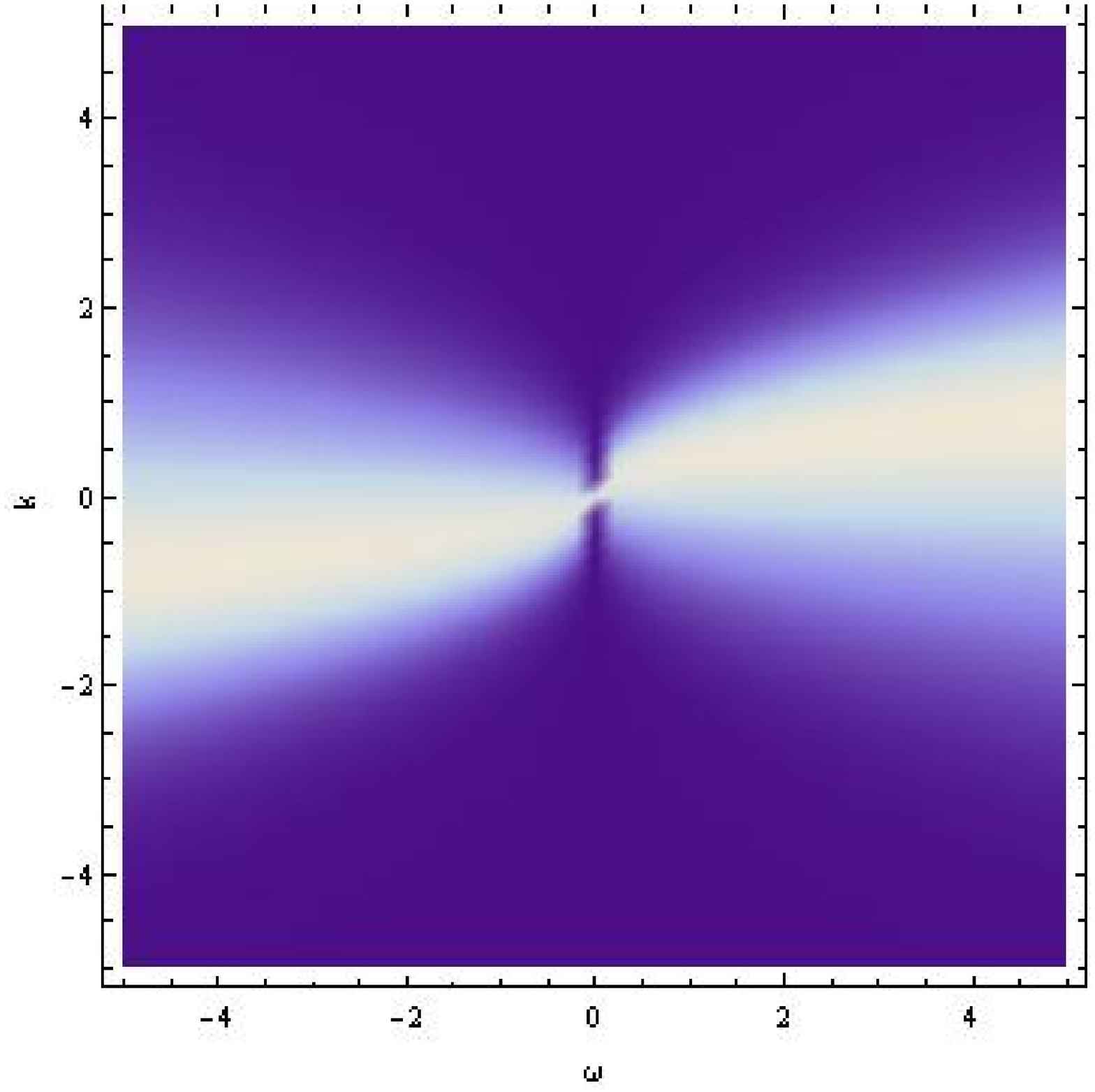}
\caption{\label{nogap} Samples of density plots of the Green
function $G_{22}$. Here we set $q=0.5$ and $m=0$. The left plot is
for $z=3$ and $\theta=1.5$ while the right one is for $z=2$ and
$\theta=1.99$.}}
\end{figure}

We showed that the variation of the hyperscaling
factor $\theta$ can generate various liquid
phases like Fermi liquids, non-Fermi liquids,
marginal-Fermi liquids and log-oscillatory. The
question is if in a holographic fermionic system
with hyperscaling violation a gap can be
dynamically generated indicating the presence of
a Mott insulating phase. In the allowed region
(\ref{ParameterRegion}) of $z$ and $\theta$, we
numerically obtained the density plot of the
Green function $G_{22}$, but we can not see the
generated gap near the zero frequency. Two
samples of our density plots are showed in Fig.
~\ref{nogap} where we have set $q=0.5$ and $m=0$.
We have checked also that other choices of $q$
and $m$  can not generate the Mott gap. It has
been firstly proposed in
\cite{1010.3238,1012.3751}  that a dipole
coupling between the fermions and gauge field
mimics doping
 in the Hubbard model and large enough dipole coupling strength  could
introduce a Mott gap phase in AdS black hole from
holography. Then, it has been addressed in
\cite{Alsup:2014uca} that the pseudo-gap phase can be
observed by studying the poles and zeros duality
through the det $G_{R}$. In order to have a
complete study of the liquid phases in
holographic Fermi systems beyond the dual AdS
geometry, it is natural to introduce the dipole
coupling in the gravity background with the
hyperscaling violation \footnote{We have
benefited  from discussions we had with Philip
Phillips on this point.}. The complete study of
the effect of the hyperscaling violation on the
liquid phase transition between Fermi Liquid,
non-Fermi liquids, pseudo-gap and Mott gap is
under preparation.

\section{Non-relativistic fermionic fixed point}
\label{secs.6}

Our studies above were focused on the dual relativistic field
theory which
 corresponds to considering the bulk action with the Lorentz covariance boundary term as
\begin{eqnarray} \label{boundary2}
S_{bdy}&=&\frac{i}{2}\int_{\partial\mathcal{M}}d^3x\sqrt{-gg^{rr}}\bar{\zeta}\zeta~.
\end{eqnarray}
In this section, taking into account that our bulk gravity is not
boost invariant, we intend to explore some properties of the
non-relativistic fixed point by  adding a Lorentz violating
boundary term into bulk action (\ref{actionspinor})
\begin{eqnarray} \label{boundary2}
S_{bdy}&=&\frac{1}{2}\int_{\partial\mathcal{M}}d^3x\sqrt{-gg^{rr}}\bar{\zeta}\Gamma^1\Gamma^2\zeta~.
\end{eqnarray}
This boundary term was first proposed in \cite{1108.1381}, where
the authors observed that the spectral function of the dual
holographic non-relativistic system showed a flat band of gapless
excitation. According to the analysis in
\cite{1108.1381,1110.4559}, the retarded Green function of
non-relativistic fixed point can be related to Green functions of
the relativistic fixed point as
\begin{eqnarray} \label{Greenfunction2}
G_{R}=\left(
\begin{array}{cc}
\frac{2G_1G_2}{G_1+G_2} & \frac{G_1-G_2}{G_1+G_2} \\
 \frac{G_1-G_2}{G_1+G_2} & \frac{-2}{G_1+G_2} \\
\end{array}
\right)
\end{eqnarray}
which is off-diagonal with $det~G_{R}=-1$ and its eigenvalue $\lambda_{\pm}$ can be expressed in terms of
$G_{I}$ as
\begin{eqnarray} \label{green2}
\lambda_{\pm}=\frac{G_{1}G_{2}-1\pm\sqrt{1+G_{1}^2+G_{2}^2+G_{1}^2G_{2}^2}}{G_{1}+G_{2}}~.
\end{eqnarray}
For simplicity, we will focus on  $m=0$ in this section. Then we
have $G_{1}=-1/G_{2}$ from the symmetry (\ref{Gsym}). Thus, the
spectral function has the form
\begin{equation}\label{spectralfunction-NR}
A_{NR}(\omega,k)=\rm Im~{Tr[G_R]}=Im~ \left[\frac{4
G_I}{1-G_{I}^2}\right]~.
\end{equation}
Thus we can obtain the spectral function of the non-relativistic
fermionic  fixed point through extracting the Green function
$G_{I}$ by solving the flow equation (\ref{DiracEF2}).

\begin{figure}
\center{
\includegraphics[scale=0.2]{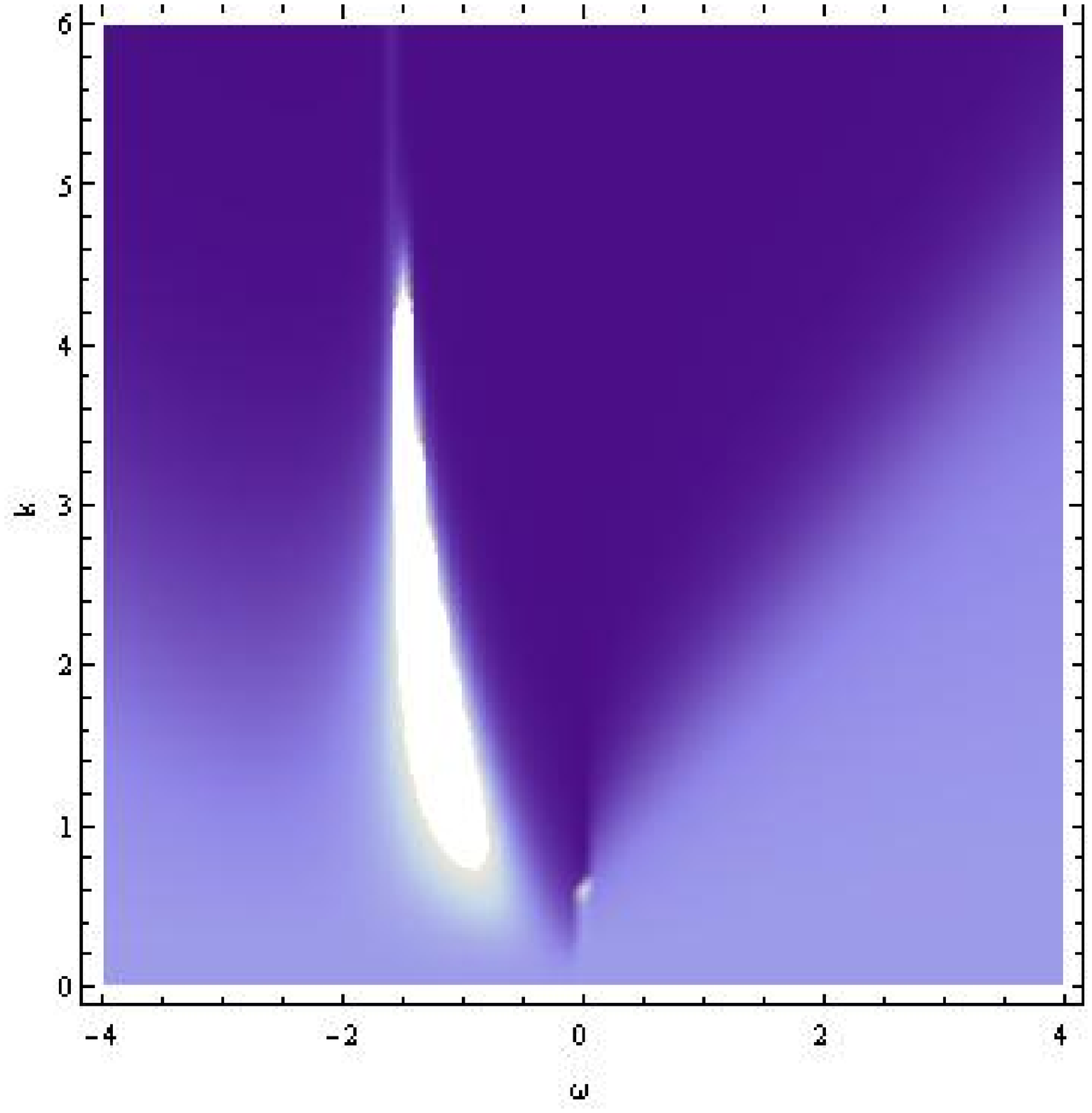}\hspace{0.5cm}
\includegraphics[scale=0.2]{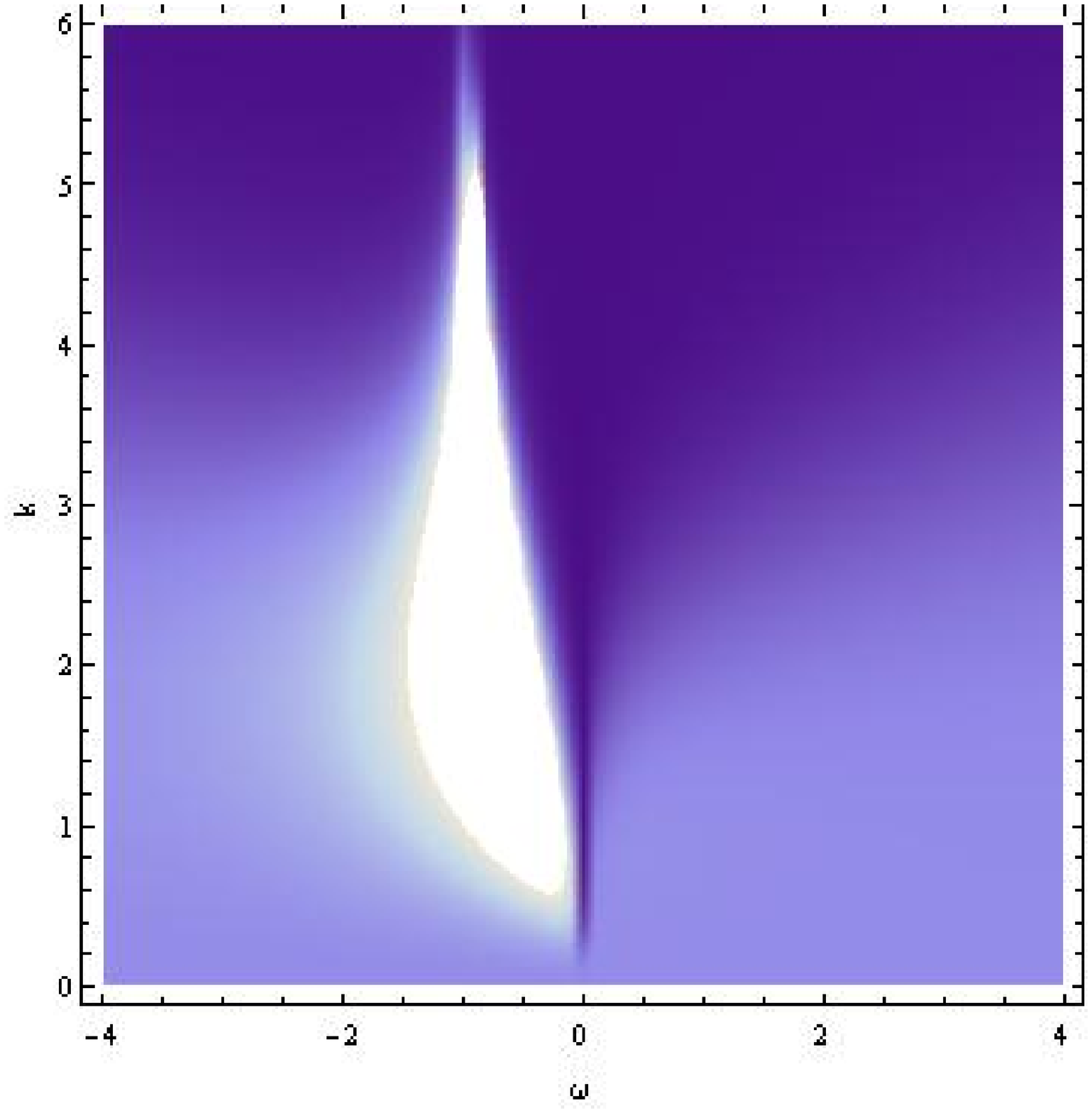}\hspace{0.5cm}
\includegraphics[scale=0.2]{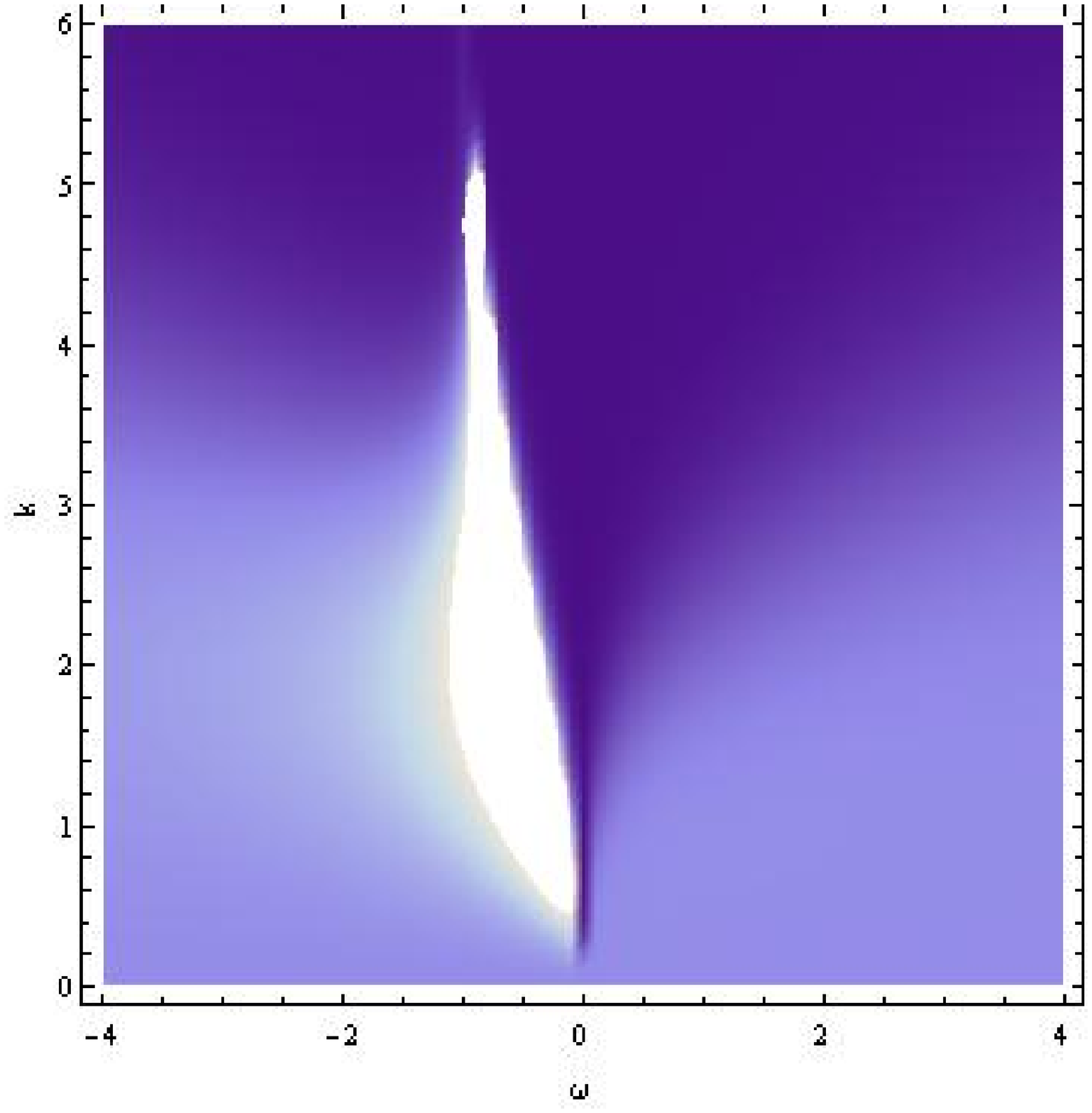}\hspace{0.5cm}
\caption{\label{fig-Anr} Density plots of the spectral function of non-relativistic fixed point.
Here we set $q=0.5$ and $m=0$. The parameters of the plots from
left to right are: $z=1; \theta=0$, $z=2; \theta=0$ and $z=2;\theta=1.5$.}}
\end{figure}
In Fig.~\ref{fig-Anr}, we show the results of the spectral
function $A_{NR}(\omega,k)$ with samples of exponents.   For
comparison, in the left plot, we reproduced the spectral function
for RN-AdS gravity. Then we turned on the dynamical exponent and
hyperscaling violation exponent. We found in hyperscaling
violation gravity, that the non-relativistic fixed point also
presented a flat band with the the similar property of poles
distributing continuously at a finite interval of momenta as that
in RN-AdS gravity. Namely, the finite band is mildly dispersed at
low momentum, but it shows strong peak at high momentum due to the
fact that the high momentum modes sit outside the lightcone and
can't decay \cite{1108.1381}.

Furthermore we observed that at large enough momentum the flat
band shifts to the frequency $\omega$ which depends on $z$ and
$\theta$. This is because the frequency $\omega$ is measured
relative to the chemical potential. The flat band corresponds to
some zero modes in the Minkowski vaccum, with the vanishing
absolute energy of Fermion  characterized by
$\omega_{eff}=\omega+q A_t$ in the Dirac equation on the boundary.
From equation~(\ref{rescalingAtmathcal2}) , it is clear that
$A_t=\mu$ at the boundary, so that the frequency $\omega$ is
related to the chemical potential with
$\mu=\frac{\sqrt{2(2-\theta)(2+z-\theta)}}{z-\theta}$ in equation
~(\ref{Q-Mu}), which is determined by both the dynamical exponent
and hyperscaling violation exponent.

We move on to explore the Fermi momentum and the effect of flat
band in the non-relativistic fixed point. Before processing, it is
necessary to point out some features of the spectral  function
(\ref{spectralfunction-NR}) with massless Dirac field. It was
found in \cite{JPWuPLBAnalytical} that in RN-AdS background the
spectral function of the non-relativistic fermionic fixed point
has the same scaling behavior at low frequency and dispersion
relation expression near the Fermi surface as the relativistic
case. Then following \cite{0907.2694,JPWuPLBAnalytical}, we can
easily obtain that in the hyperscaling violation gravity, the
scaling of $A_{NR}(\omega,k)$ near small $\omega$ is the same as
the relativistic $A(\omega,k)$ in
equation~(\ref{spectralfunction}) and the dispersion relation of
$A_{NR}(\omega,k)$ coincides with equation~(\ref{LdispersionA}).
Thus, we can numerically determine the Fermi momentum and employ
Eq.~(\ref{LdispersionA}) to get the dispersion relation of the
non-relativistic fixed point.

We give the values of the Fermi momentum with $m=0$ and $q=0.5$ in
Table~\ref{tablekkf}.  To compare, we set the same exponents for
the non-relativistic and relativistic cases and we have chosen $z$
and $\theta$ as that in Fig.~\ref{G22vkq05z1p02m0}. From the
table, we see that similar to the relativistic case, the Fermi
momentum for non-relativistic case also increases as the
hyperscaling violation exponent. However, with any $z$ and
$\theta$, the Fermi momentum for non-relativistic case is lower
than that in the relativistic case, meaning that the Fermi surface is
suppressed by the flat band. This phenomenon was also observed in
RN-AdS gravity \cite{1110.4559} and charged
 dilaton gravity \cite{1210.5735}. It would be interesting to consider the
case with massive fermions.
\begin{table}
\centering
\begin{tabular}{|c|c|c|c|c|c|}
  \hline
  $\theta$ & 0 &0.1 & 0.2 &0.3  &0.4 \\ \hline
  $k_F(R)$ &0.681 &0.688& 0.694& 0.701 & 0.708\\ \hline
  $k_F(NR)$ &0.44  & 0.456 & 0.47 &0.485&0.498  \\ \hline
\end{tabular}
\caption{The Fermi momentum changes with the hyperscaling exponent
with fixed dynamical  exponent $z=1.2$ for relativistic and
non-relativistic fixed point.} \label{tablekkf}
\end{table}

\section{Conclusions and discussion}
\label{secs.7}

We have studied the features of the fermionic response in a
holographic system with a charged black hole with hyperscaling
violation in the bulk.  Since the near horizon geometry is
$AdS_2$, we followed the matching method of \cite{0907.2694} to
obtain the analytical expressions of the UV Green's function and
the dispersion relation, which have a similar form as that in
RN-AdS black hole \cite{0907.2694} and charged Lifshitz background
 \cite{wu-lifshitz}. However, since both  the Fermi momentum
$k_F$ and the dimension $\delta_k$ in the IR CFT depends on the
Lifshitz dynamical critical exponent and hyperscaling violating
exponent, the dispersion relation also varies with the two
exponents.

We numerically determined the Fermi momentum
$k_F$ to obtain a relation  between the scaling
exponent $\delta$ of the dispersion relation and
the hyperscaling violating exponent $\theta$. We
found that for the case of fixed $q$, $z$ and
zero mass of the fermions, the exponent $\delta$
decreases rapidly as the hyperscaling exponent
becomes larger. This indicates that with the
increase of $\theta$, the degree of deviation
from the Landau Fermi liquid becomes smaller.

Since the boundary Green's function  depends on
the mass of the fermions, we studied the  mass
dependence. We found that for small Lifshitz
exponent $z$ and hyperscaling exponent $\theta$,
with the increase of $m$, the Fermi momentum
$k_F$ almost linearly decreases as that in RN-AdS
black hole and the scaling exponent $\delta$
decreases. We found that there is a transition
from non-Fermi liquid to Fermi liquid as the $m$
decreases. For large $z$ and $\theta$, the Fermi
momentum $k_F$ still linearly decreases with the
increase of $m$ in the region of large $m$, but
this does not hold when $m$ approaches the low
bound ($m=-0.5$). Also, in the region of $m\in
(-0.5,0.5)$, the scaling exponent $\delta$ is
always larger than $1$, which indicates that it
is always in a non-Fermi liquid phase.

We looked for the possibility to generate
dynamically a gap indicating the presence of a
Mott insulating phase. In the allowed region  of
$z$ and $\theta$ values we numerically obtained
the density plot of the Green function $G_{22}$,
but we failed to observe a generation of a  gap
near the zero frequency.  We have also checked
that other choices of values of $q$ and $m$ could
not generate the Mott gap. This indicates that in
order to generate a Mott gap we need to introduce
another scale in the fermionic system, such as a
dipole moment.

Finally, we added a Lorentz-violating boundary term into the bulk
action and investigated the holographic non-relativistic fixed
point dual to the hyperscaling violation gravity bulk. Similar to
the case in RN-AdS black hole \cite{1108.1381}, we also observed a
flat band of gapless excitation which suppressed the Fermi
momentum. Furthermore, here the spectral function at high momentum
shifted to frequency dependent on the dynamical exponent and
hyperscaling violation exponent, because in the flat band state,
the frequency is measured relative to the chemical potential
determined by the two exponents in the form of equation
(\ref{Q-Mu}).

It would be interesting  to extend this study
into a system with a dipole coupling  between the
gauge field and the fermions and see how the
hyperscaling violation imprints  on the
generation of an insulating phase both in the relativistic and non-relativistic fixed points. Another
interesting direction would be to study the case of non-zero temperature. In this way we can study how
the phenomena and features disclosed due to the
presence of  hyperscaling violation would
complement the behaviour of quantum liquids and
their realization in condensed matter physics.

\begin{acknowledgments}
We thank Philip Phillips for valuable
discussions. X. M. K. and E. P. are supported by
ARISTEIA II action of the operational programme
education and long life learning which is
co-funded by the European Union (European Social
Fund) and National Resources. B. W. and J. P. W. are
supported by the Natural Science Foundation of China.
J. P. Wu is also supported by Program for Liaoning Excellent Talents in University (No. LJQ2014123).
\end{acknowledgments}

\end{document}